\documentclass[12pt,journal,a4paper,onecolumn]{IEEEtran}
\usepackage{graphicx}
\usepackage{amsmath,epsfig,amssymb, amstext, verbatim,amsopn,cite,subfigure,multirow,multicol,lipsum}
\usepackage{balance}
\usepackage{url}
\usepackage{amsfonts}
\usepackage{epsfig}
\usepackage{slashbox}
\usepackage{setspace}
\usepackage{stmaryrd}
\usepackage{psfrag}	
\usepackage{multirow}
\usepackage{float}
\usepackage{slashbox}
\usepackage[process=auto]{pstool}
\usepackage{etoolbox}
\allowdisplaybreaks
\newtoggle{OneColumn}

\toggletrue{OneColumn}
\iftoggle{OneColumn}{%
  \doublespacing
}{%
}

\newcommand{\Equal}{\hspace{-0.5mm}=\hspace{-0.5mm}}
\newcommand{\Add}{\hspace{-0.5mm}+\hspace{-0.5mm}}
\newcommand{\Minus}{\hspace{-0.5mm}-\hspace{-0.5mm}}

\newcommand{\Lequal}{\hspace{-0.5mm}\leq\hspace{-0.5mm}}
\newcommand{\Gequal}{\hspace{-0.5mm}\geq\hspace{-0.5mm}}

\newtheorem{theo}{Theorem}

\newtheorem{remk}{Remark}

\newtheorem{Corol}{Corollary}

\EndPreamble
\begin{document}
%\begin{doublespace}

\title{Adaptive Mode Selection for Bidirectional Relay Networks -- Fixed Rate Transmission}
\author{Vahid Jamali$^\dag$, Nikola Zlatanov$^\ddag$, and Robert Schober$^\dag$ \\
\IEEEauthorblockA{$^\dag$ Friedrich-Alexander-University Erlangen-N\"{u}rnberg (FAU), Germany \\
 $^\ddag$ University of British Columbia (UBC), Vancouver, Canada}
}

\maketitle

\begin{abstract}
In this paper, we consider the problem of sum throughput maximization for  bidirectional relay networks with block fading. Thereby, user 1 and user 2 exchange information only via a relay node, i.e., a direct link between both users is not present. We assume that channel state information at the transmitter (CSIT) is not available and/or only one coding and modulation scheme is used at the transmitters due to complexity constraints. Thus, the nodes transmit with a fixed predefined rate regardless of the channel state information (CSI).  In general, the nodes in the network can assume one of  three possible states in each time slot, namely the transmit, receive, and silent state. Most of the existing protocols assume a fixed schedule for the sequence of the states of the nodes.  In this paper, we abandon the restriction of having a fixed and predefined schedule and propose a new protocol which, based on the CSI at the receiver (CSIR), selects the optimal states of the nodes in each time slot such that the sum  throughput  is maximized.  To this end, the relay has to be equipped with two buffers for  storage of the information received from the two users. Numerical results show that the proposed protocol  significantly outperforms the existing protocols.
\end{abstract}

%\begin{keywords}
%Bidirectional transmission, sum throughput, outage probability, adaptive mode selection,  buffer-aided relay.
%\end{keywords}

\section{Introduction} \label{Sec I (Intro)}

Relaying in cooperative  communication was originally proposed as a means to extend the system coverage and to increase the throughput and reliability of wireless networks \cite{Meulen}. Recently, bidirectional relaying, where two users exchange information via a relay node, has attracted much interest. In particular, this simple network architecture can be used to model several practical applications such as satellite communication and cellular communication via a base
station.  

Several protocols have been
proposed for the bidirectional relay network under the practical half-duplex
constraint,  i.e.,  a  node  cannot  transmit  and  receive  at  the
same  time  and  in  the  same  frequency  band.  The  simplest
protocol is the traditional two-way relaying protocol in which the bidirectional transmission is accomplished using four successive point-to-point phases: user 1-to-relay, relay-to-user 2, user 2-to-relay, and relay-to-user 1. However, this protocol suffers from a loss in spectral efficiency due to the pre-log factor of $\frac{1}{2}$ caused by the two-hop transmission architecture. To increase spectral efficiency of bidirectional relaying, the time division broadcast
(TDBC)  protocol was proposed in \cite{TDBC} which combines the relay-to-user 1 and relay-to-user  2  phases  into  one  phase,  the  broadcast  phase. Thereby, the relay transmits to both users simultaneously.
 To further enhance spectral efficiency, the multiple access
broadcast (MABC) protocol was proposed in \cite{MABC} where the user 1-to-relay and
user 2-to-relay phases are also combined into one phase, the
multiple-access phase. In the multiple-access phase, both
users simultaneously transmit to the relay.  A significant research effort has been dedicated
to  obtaining  the  achievable rate  region  of  the  bidirectional
relay  channel \cite{Tarokh,BocheIT,TDBC,MABC}.  However, these rate regions were derived for adaptive rate transmission which requires the availability of  channel state information at the transmitter (CSIT) and the capability of using appropriate coding and modulation schemes such that the transmitters can adapt their transmission rates to the channel capacity. For the cases when CSIT is not available and/or only one coding and modulation scheme is used, protocols assuming adaptive rate transmission are not applicable.  Instead, the transmitters have to transmit with fixed rates regardless of the channel state information (CSI) of the involved links. For fixed rate transmission, not the achievable rate region but other performance metrics  such as throughput and outage probability have to be considered \cite{Outage2,Multiplexing,Outage1}. 

\begin{table}
\label{Modes}
\caption{Relevant Transmission Modes in the Considered Bidirectional Relay Networks (T: Transmit, R: Receive, S: Silent).} 
\begin{center}
\scalebox{0.8}{
\begin{tabular}{|| c | c | c | c |c | c | c | c ||}
  \hline                  
\textbf{Transmission Mode} & $\mathcal{M}_1$ & $\mathcal{M}_2$ & $\mathcal{M}_3$ & $\mathcal{M}_4$ & $\mathcal{M}_5$ & $\mathcal{M}_6$ & $\mathcal{M}_7$ \\ \hline
\textbf{User 1} & T & S & T & R & S & R & S \\ \hline
\textbf{User 2} & S & T & T & S & R & R & S \\ \hline
\textbf{Relay} & R & R & R & T & T & T & S \\ \hline
  
\end{tabular}
}
\end{center}
\vspace{-0.5cm}
\end{table}
 
In general, the nodes in the network can assume one of   three possible states in each time slot, namely the transmit,  receive, and silent state. Among the $3^3=27$ possible combinations of the states of the nodes, only seven combinations  are relevant in the considered bidirectional relay network due to the half-duplex constraint,  see Table I. Each of these combinations is referred to as a transmission mode.  The seven relevant transmission modes are given in the following: four point-to-point modes (user 1-to-relay, user 2-to-relay, relay-to-user 1, relay-to-user 2), a multiple access mode (both users to the relay), a broadcast mode (the relay to both users), and a silent mode (all nodes are silent). Previously  proposed protocols utilize a fixed and predefined schedule of using a subset of the available transmission modes \cite{Meulen,Tarokh,TDBC,MABC,BocheIT,Outage2,Multiplexing,Outage1}. 
However, for one-way relaying, it was shown in \cite{NikolaJSAC} and \cite{NikolaMixed} that a considerable gain is obtained with adaptive link selection where based on the CSI either the source-relay or relay-destination links are selected for transmission in each time slot. In particular,  optimal link selection policies achieving the maximum throughput of one-way relay networks were derived for adaptive and fixed rate transmission in \cite{NikolaJSAC}  and \cite{NikolaMixed}, respectively. For bidirectional relay networks, the optimal mode selection policy achieving the maximum sum rate was derived  in \cite{GlobeComIEEE}  for adaptive rate transmission. A simpler protocol was proposed in \cite{PopovskiLetter} where only two point-to-point modes and the broadcast mode were available for selection. Thus, this protocol has a lower performance compared to the protocol in \cite{GlobeComIEEE}.   

Motivated by the performance gains reported in \cite{NikolaJSAC,NikolaMixed,GlobeComIEEE,PopovskiLetter}, in this paper, we consider the problem of sum throughput maximization for the bidirectional relay network under the constraint of fixed rate transmission. In particular, we propose a protocol which is not restricted to have a fixed and predefined schedule of using the available transmission modes. Instead, based on the channel state information at the receiver (CSIR), the optimal transmission mode is selected in each time slot such that the sum throughput is maximized. To this end, the relay has to be equipped with two buffers for storage of the information received from user 1 and user 2. Considering fixed rate transmission is of both practical and theoretical  interest. Specifically, simpler transmitters  can be employed for fixed rate transmission than for adaptive rate transmission since only one coding and modulation scheme is needed. Moreover, the overhead required for feedback information is very low as only three bits of feedback are required in the proposed protocol to select one of the seven possible transmission modes.  From a theoretical point of view, it is interesting to obtain  performance limits for bidirectional relay networks under the constraint of a fixed transmission rate.

\section{System Model}\label{SysMod}
In this section, we describe the system model and analyze the seven possible transmission modes used to develop the proposed protocol. 

\subsection{Channel Model}
We consider a simple network in which user 1 and user 2
exchange information with the help of a relay node as shown
in Fig. 1. We assume that there is no direct link between the users, and thus, user 1 and user 2 communicate with
each other only through the relay node. We assume that all
three nodes in the network are half-duplex. Furthermore, we
assume that time is divided into slots of equal length and 
each node transmits codewords which span one time slot. We assume
that the user-to-relay and relay-to-user channels are impaired
by additive white Gaussian noise (AWGN) and block fading, i.e., the channel 
coefficients are constant during one time slot and change from
one time slot to the next. Moreover, in each time slot, the
channels are assumed to be reciprocal, i.e., the
user 1-to-relay and the user 2-to-relay channels are identical to
the relay-to-user 1 and relay-to-user 2 channels, respectively. The channel reciprocity assumption is valid
for time-division-duplex (TDD) systems where the user-to-relay and relay-to-user links utilize the same
frequency band. Let $h_1(i)$ and $h_2(i)$ denote the channel fading coefficients between user 1 and the relay and between user 2 and the relay in the $i$-th time slot, respectively. Fading gains $|h_1(i)|^2$ and $|h_2(i)|^2$ are assumed to be ergodic and stationary random processes with means $\Omega_1=E\{|h_1(i)|^2\}$ and  $\Omega_2=E\{|h_2(i)|^2\}$, respectively, where $E\{\cdot\}$ denotes expectation.  
Furthermore, $\gamma_1(i)=\gamma|h_1(i)|^2$ and $\gamma_2(i)=\gamma|h_2(i)|^2$ denote the instantaneous signal-to-noise ratios (SNRs) of the links between user 1 and the relay and user 2 and the relay, respectively, where $\gamma=\frac{P}{\sigma_n^2}$ is the transmit SNR of the nodes, $P$ is the transmit power of the nodes, and $\sigma_n^2$ is the noise variance at the receivers.  Since the noise is AWGN, we assume that  the transmitted codewords of user 1, user 2, and the relay are comprised of
symbols which are Gaussian distributed random variables with variance $P$. We also assume that all nodes transmit with fixed  rate $R_0$.

%%%%%%%%%%%%%%%%%%%%%%%%%%%%%%%%%%%%%%%%%%%%%% Figure %%%%%%%%%%%%%%%%%%%%%%%%%%%%%%%%%%%%%%%%%
\begin{figure}
\centering
\pstool[width=0.5\linewidth]{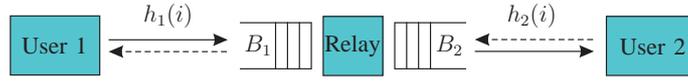}{
\psfrag{U1}[c][c][0.75]{$\text{User 1}$}
\psfrag{U2}[c][c][0.75]{$\text{User 2}$}
\psfrag{R}[c][c][0.75]{$\text{Relay}$}
\psfrag{h1}[c][c][0.75]{$h_1(i)$}
\psfrag{h2}[c][c][0.75]{$h_2(i)$}
\psfrag{B1}[c][c][0.75]{$B_1$}
\psfrag{B2}[c][c][0.75]{$B_2$}}
\caption{Bidirectional relay network consisting of two users and a buffer-aided relay.}
\label{FigSysMod}
\vspace{-0.4cm}
\end{figure}
%%%%%%%%%%%%%%%%%%%%%%%%%%%%%%%%%%%%%%%%%%%%%%%%%%%%%%%%%%%%%%%%%%%%%%%%%%%%%%%%%%%%%%%%%%%%%%%

\subsection{Analysis of the Transmission Modes}

In the considered bidirectional relay channel, only seven transmission modes are relevant, cf. Table I. The transmission modes are denoted by  ${\cal M}_1,...,{\cal M}_7$. In order to avoid information loss, we select  transmission modes ${\cal M}_1,...,{\cal M}_6$ only if the information can be decoded successfully at the receiver(s).  Otherwise, we select silent mode $\mathcal{M}_7$. Let $B_1$ and $B_2$ denote two infinite-size buffers at the relay which store the  information received from user 1 and user 2, respectively. Moreover, $Q_j(i), \,\, j\in\{1,2\}$, denotes the amount of normalized information in bits/symbol available in buffer $B_j$ in the $i$-th time slot. Using these notations and assumptions, the transmission modes and the dynamics of the queues at the buffers are presented in the following:

${\cal M}_1$: User 1 transmits to the relay and user 2 is silent. For this mode, the relay can decode the information successfully only if $\gamma_1(i)>\gamma_\mathrm{thr}$ holds, where $\gamma_\mathrm{thr}=2^{R_0}-1$. Thereby, the relay stores the information in buffer $B_1$ and the amount of information in buffer $B_1$ increases to $Q_1(i)=Q_1(i-1)+R_{0}$.

${\cal M}_2$: User 2 transmits to the relay and user 1 is silent. For this mode, the relay can decode the information successfully only if $\gamma_2(i)>\gamma_\mathrm{thr}$ holds.  Thereby, the relay stores the information in buffer $B_2$ and the amount of information in buffer $B_2$ increases to $Q_2(i)=Q_2(i-1)+R_{0}$.

${\cal M}_3$: Both users 1 and 2 transmit to the relay  simultaneously. For this mode, we assume that  multiple access transmission is used, see \cite{Cover}. The relay can decode the information from both users successfully only if $\gamma_1(i)>\gamma_\mathrm{thr}$, $\gamma_2(i)>\gamma_\mathrm{thr}$, and $\gamma_1(i)+\gamma_2(i)>\gamma_\mathrm{thr}^\mathrm{sum}$ hold, where $\gamma_\mathrm{thr}^\mathrm{sum} = 2^{2R_0}-1$.  Thereby, the relay stores the information received from user 1 and user 2 in buffers $B_1$ and $B_2$, respectively. Therefore, the amounts of information in buffers $B_1$ and $B_2$ increase to $Q_1(i)=Q_1(i-1)+R_{0} $ and $Q_2(i)=Q_2(i-1)+R_{0}$, respectively.

${\cal M}_4$: The relay transmits the information received  from user 2 to user 1. Specifically, the relay extracts $R_0$ bits of information from buffer $B_2$, encodes it into a codeword, and transmits it to user 1. Thus, a successful transmission for this mode depends on both availability of information in buffer $B_2$ and condition of the relay-to-user 1 link. In particular, user 1 can decode $R_0$ bits of information successfully in this mode only if $\gamma_1(i)>\gamma_\mathrm{thr}$ and $Q_2(i-1) \geq R_0$ hold.  Thereby,  the amount of information in buffer $B_2$ decreases to $Q_2(i)\Equal Q_2(i- 1)\Minus R_{0}$.

${\cal M}_5$: This mode is identical to ${\cal M}_4$ with user 1 and 2 switching roles. For this mode, user 2 can decode $R_0$ bits of information successfully only if $\gamma_2(i)>\gamma_\mathrm{thr}$ and $Q_1(i-1) \geq R_0$ hold.  Thereby, the amount of information in buffer $B_1$ decreases to $Q_1(i)\Equal Q_1(i\Minus 1)\Minus R_{0}$.

${\cal M}_6$: The relay broadcasts to both user 1 and user 2 the information received from user 2 and user 1, respectively. Specifically, the relay extracts $R_0$ bits of information intended for user 2 from buffer $B_1$ and  $R_0$ bits of information intended for user 1 from buffer $B_2$. Then, based on the scheme in \cite{BocheIT}, it constructs a superimposed codeword which contains the information of both users and broadcasts it to the users. For this mode, both users can decode $R_0$ bits of information successfully only if  $\gamma_1(i)>\gamma_\mathrm{thr}$, $\gamma_2(i)>\gamma_\mathrm{thr}$, $Q_1(i-1)\geq R_0$, and $Q_2(i-1)\geq R_0$ hold. Thereby, the amounts of information in buffers $B_1$ and $B_2$ decrease to $Q_1(i)\Equal Q_1(i\Minus 1)\Minus R_{0}$ and $Q_2(i)\Equal Q_2(i\Minus 1)\Minus R_{0}$, respectively.

${\cal M}_7$: For this mode all nodes are silent and the status of the queues at the buffers does not change. 

\subsection{Mode Selection Variables and SNR Regions}

We introduce seven binary variables,  $q_k(i) \in\{0,1\}, \,\,k=1,...,7$, where  $q_k(i)$ indicates whether or not transmission mode  $\mathcal{M}_k$ is selected in the $i$-th time slot. In particular, $q_k(i)=1$ if  mode $\mathcal{M}_k$ is selected and $q_k(i)=0$ if it is not selected in the $i$-th time slot. Furthermore, since in each time slot  only one of the seven transmission modes can be selected, only one of the mode selection variables is equal to one and the others are zero, i.e., $\mathop \sum_{k = 1}^7 q_k(i)=1$ holds. 

Moreover, let $O_k(i)\in\{0,1\}$ be a binary variable specifying the decodability of information at the receivers for mode $\mathcal{M}_k$ based on CSI in the $i$-th time slot. In particular, assuming the availability of information at the transmitters, $O_k(i)=1$ if the transmitted information can be successfully decoded at the receiver(s) for mode $\mathcal{M}_k$ and $O_k(i)=0$ if the transmitted information cannot be successfully decoded at all receivers for mode $\mathcal{M}_k$ in the $i$-th time slot. Fig. \ref{FigOutReg} illustrates five possible regions for the instantaneous link SNRs based on the decodability of information at the receivers, i.e., $\boldsymbol{\gamma}(i)=[\gamma_1(i),\gamma_2(i)]\in\mathcal{R}_l,\,\,l=1,\dots,5$. In particular, in region $\mathcal{R}_1$,  for all the modes, the receivers can decode the information.  In region $\mathcal{R}_2$, only the relay cannot decode both users' information in the multiple access mode $\mathcal{M}_3$. In regions $\mathcal{R}_3$ and $\mathcal{R}_4$, only the receivers for modes $\{\mathcal{M}_1,\mathcal{M}_4\}$ and $\{\mathcal{M}_2,\mathcal{M}_5\}$ can successfully decode the information, respectively. Finally, in region $\mathcal{R}_5$, none of the receivers for modes ${\cal M}_1,...,{\cal M}_6$ can decode the information successfully. We also define $P_{\mathcal{R}_l}=\Pr\{\boldsymbol{\gamma}(i) \in \mathcal{R}_l\}$ for future reference.
%%%%%%%%%%%%%%%%%%%%%%%%%%%%%%%%%%%%%%%%%%%%%%%%%%%%%%%%%%%%%%%%%%%%%%%%%%%%%%%%%%%%%%%%%%%%%%%%%%%%%

\section{Adaptive Mode Selection}

In this section, we first present the problem formulation for sum throughput maximization with adaptive mode selection. Then, we propose a protocol for the optimal mode selection policy as a solution of the optimization problem. Moreover, we analyze the performance of the proposed protocol in the high SNR regime. 

\subsection{Problem Formulation}

%%%%%%%%%%%%%%%%%%%%%%%%%%%%%%%%%%%%%%%%%%%%%%%% Figure %%%%%%%%%%%%%%%%%%%%%%%%%%%%%%%%%%%%%%%%%
\begin{figure}
\centering
    \pstool[width=0.4\linewidth]{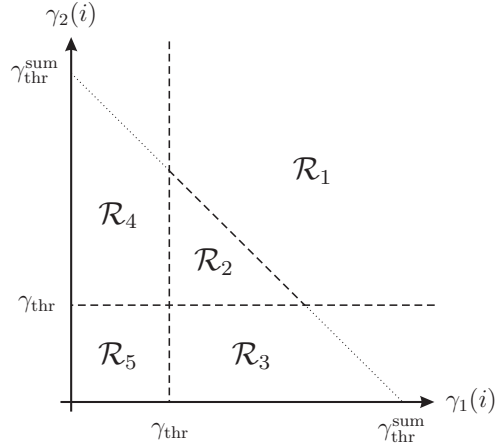}{
\psfrag{R1}[c][c][1]{$\mathcal{R}_1$}
\psfrag{R2}[c][c][1]{$\mathcal{R}_2$}
\psfrag{R3}[c][c][1]{$\mathcal{R}_3$}
\psfrag{R4}[c][c][1]{$\mathcal{R}_4$}
\psfrag{R5}[c][c][1]{$\mathcal{R}_5$}
\psfrag{G1}[c][c][0.8]{$\gamma_1(i)$}
\psfrag{G2}[c][c][0.8]{$\gamma_2(i)$}
\psfrag{Gthr}[c][c][0.8]{$\gamma_{\mathrm{thr}}$}
\psfrag{Gsum}[c][c][0.8]{$\gamma_{\mathrm{thr}}^{\mathrm{sum}}$}}
\caption{Instantaneous SNR regions based on the decodability of information at the receivers in the different transmission modes where $\gamma_{\mathrm{thr}}=2^{R_0}-1$ and $\gamma_{\mathrm{thr}}^{\mathrm{sum}} = 2^{2R_0}-1$.}
\label{FigOutReg}
\vspace{-0.4cm}
\end{figure}
%%%%%%%%%%%%%%%%%%%%%%%%%%%%%%%%%%%%%%%%%%%%%%%%%%%%%%%%%%%%%%%%%%%%%%%%%%%%%%%%%%%%%%%%%%%%%%%%%

 Let $\bar{R}_{12}$ and $\bar{R}_{21}$ denote the average throughputs achieved for the user 1-to-user 2 and user 2-to-user 1 transmissions, respectively. In this paper, our goal is to devise a protocol that optimally selects the transmission mode based on the instantaneous CSI in each time slot such that the sum throughput, i.e., $\bar{R}_{\mathrm{sum}}=\bar{R}_{12}+\bar{R}_{21}$, is maximized. On the other hand, the outage probability for each transmission direction is defined as the reduction of the achievable throughput compared to the maximum throughput when the receivers can always decode the information \cite{NikolaMixed}. Mathematically, we can write the outage probabilities for both transmission directions as
%%%%%%%%%%%%%%%%%%%%%%%%%%%%%%%%%%%%%%%%%
\begin{IEEEeqnarray}{lll}\label{SysOut}
F^{\mathrm{out}}_{12} = 1- \frac{\bar{R}_{12}}{\bar{R}_{12}^{\max}} ,\quad F^{\mathrm{out}}_{21} = 1- \frac{\bar{R}_{21}}{\bar{R}_{21}^{\max}}
\end{IEEEeqnarray}
 %%%%%%%%%%%%%%%%%%%%%%%%%%%%%%%%%%%%%%%%%%%%%%%%%%%%%%%%%%%%%%%%%%%%%%%%%%%%%%
where $\bar{R}_{12}^{\max}=\bar{R}_{21}^{\max}=R_0/2$ \cite{NikolaMixed}. Therefore, by maximizing the sum throughput, the sum of the individual outage probabilities is minimized. In other words, the  outage probability of the system is minimized, i.e., $F^{\mathrm{out}}_{\mathrm{sys}} = \frac{F^{\mathrm{out}}_{12} + F^{\mathrm{out}}_{21}}{2}$.

We assume that user 1 and user 2 always have enough information to send in all time slots and that the number of time slots, $N$, satisfies $N\to \infty$. Moreover, the average throughputs of the user 1-to-relay, user 2-to-relay, relay-to-user 1, and relay-to-user 2 links are denoted by $\bar{R}_{1r}$, $\bar{R}_{2r}$, $\bar{R}_{r1}$, and $\bar{R}_{r2}$, respectively, and are given by
%%%%%%%%%%%%%%%%%%%%%%%%%%%%%%%%%%%%%%%%%
\begin{IEEEeqnarray}{lll}\label{RatReg123}
    \bar{R}_{1r} &= \underset{N\to \infty}{\lim} \frac{1}{N}\mathop \sum \limits_{i = 1}^N \left[ q_1(i)O_1(i)+q_3(i)O_3(i)\right] R_{0} \IEEEyesnumber\IEEEyessubnumber \\
		\bar{R}_{2r} &=  \underset{N\to \infty}{\lim} \frac{1}{N}\mathop \sum \limits_{i = 1}^N \left[ q_2(i)O_2(i)+q_3(i)O_3(i)\right] R_{0} \IEEEyessubnumber\\
    \bar{R}_{r1}  &\hspace{-0.03cm}=\hspace{-0.2cm} \underset{N\to \infty}{\lim}\frac{1}{N}\hspace{-0.1cm} \mathop \sum \limits_{i = 1}^N \hspace{-0.1cm} \left[ q_4(i)O_4(i)\hspace{-0.1cm}+\hspace{-0.1cm} q_6(i)O_6(i)\right]\hspace{-0.1cm} \min\{\hspace{-0.05cm} R_{0},\hspace{-0.05cm}Q_{2}(i\hspace{-0.1cm}-\hspace{-0.1cm} 1)\}\,\, \IEEEeqnarraynumspace\IEEEyessubnumber \\
		\bar{R}_{r2}  &\hspace{-0.03cm}=\hspace{-0.2cm} \underset{N\to \infty}{\lim}\frac{1}{N}\hspace{-0.1cm} \mathop \sum \limits_{i = 1}^N \hspace{-0.1cm} \left[ q_5(i)O_5(i)\hspace{-0.1cm}+\hspace{-0.1cm} q_6(i)O_6(i)\right]\hspace{-0.1cm} \min\{\hspace{-0.05cm} R_{0},\hspace{-0.05cm}Q_{1}(i\hspace{-0.1cm}-\hspace{-0.1cm} 1)\} \,\,\IEEEeqnarraynumspace \IEEEyessubnumber
\end{IEEEeqnarray}
%%%%%%%%%%%%%%%%%%%%%%%%%%%%%%%%%%%%%%%%%%%%%%%%%%%%%%%%%%%%%%%%%%%%%%%%%%%%%%%%%%%%%%%%%%%%%
Furthermore, the average information received at user 2 from user 1  is identical to the average information that user 2  receives from the relay, i.e.,  $\bar{R}_{12}=\bar{R}_{r2}$. Similarly, we obtain that  $\bar{R}_{21}=\bar{R}_{r1}$ has to hold. 

In \cite[Theorem 1]{GlobeComIEEE}, we have introduced a useful condition for the queues of the buffers at the relay for adaptive rate transmission. The same condition must hold when the nodes transmit with fixed rates. In particular, the queues of the buffers $B_1$ and $B_2$ at the relay have to be at the edge of non-absorbtion. More precisely, for the maximum throughput, $\bar{R}_{1r}=\bar{R}_{r2}$ and $\bar{R}_{2r}=\bar{R}_{r1}$ must hold where $\bar{R}_{r1}$ and $\bar{R}_{r2}$ have to satisfy
%%%%%%%%%%%%%%%%%%%%%%%%%%%%%%%%%%%%%%%%%
\begin{IEEEeqnarray}{lll}\label{RatRegApp456-buffer}
\bar{R}_{r1}= \underset{N\to \infty}{\lim}\frac{1}{N}\mathop \sum \limits_{i = 1}^N \left[ q_4(i)O_4(i)+q_6(i)O_6(i)\right] R_{0} \IEEEyesnumber\IEEEeqnarraynumspace \IEEEyessubnumber \\
\bar{R}_{r2} = \underset{N\to \infty}{\lim}\frac{1}{N}\mathop \sum \limits_{i = 1}^N \left[ q_5(i)O_5(i)+q_6(i)O_6(i)\right] R_{0} \IEEEeqnarraynumspace\IEEEyessubnumber
\end{IEEEeqnarray}
 %%%%%%%%%%%%%%%%%%%%%%%%%%%%%%%%%%%%%%%%%%%%%%%%%%%%%%%%%%%%%%%%%%%%%%%%%%%%%%
For the proof please refer to \cite{GlobeComIEEE}. In particular, the capacity terms for adaptive rate transmission  in \cite{GlobeComIEEE} have to be replaced by $O_k(i)R_0,\,\,\forall i,k$ for fixed rate transmission. We note that with the above conditions, the effect of the queues at the relay becomes negligible for the maximum sum throughput, i.e., the relay  always has information to transmit and thus, (\ref{RatReg123}c) and (\ref{RatReg123}d) simplify to (\ref{RatRegApp456-buffer}a) and (\ref{RatRegApp456-buffer}b), respectively. 

Now, we are ready to present the considered optimization problem. The sum throughput maximization problem with adaptive mode selection is formulated as follows
%%%%%%%%%%%%%%%%%%%%%%%%%%%%%%%%%%%%%%%%%%%%%%%% EQUATION %%%%%%%%%%%%%%%%%%%%%%%%%%%%%%%%%%%%%%%%%
\begin{IEEEeqnarray}{Cll}\label{Prob}
    {\underset{q_k(i)\,\, \forall i,k}{\mathrm{maximize}}}\,\, & \bar{R}_{\mathrm{sum}} \nonumber \\
    \mathrm{subject\,\, to} \,\, &\mathrm{C1}:\quad \bar{R}_{1r}=\bar{R}_{r2}  \nonumber \\
    &\mathrm{C2}:\quad \bar{R}_{2r}=\bar{R}_{r1} \nonumber \\
		&\mathrm{C3}:\quad \sum\limits_{k = 1}^7 {q_k}\left( i \right) = 1, \,\, \forall i   \nonumber \\
    &\mathrm{C4}:\quad q_k(i)\in\{0,1\}, \,\, \forall i, k 
\end{IEEEeqnarray}
%%%%%%%%%%%%%%%%%%%%%%%%%%%%%%%%%%%%%%%%%%%%%%%%%%%%%%%%%%%%%%%%%%%%%%%%%%%%%%%%%%%%%%%%%%%%%%%%%%%%%
where constraints $\mathrm{C1}$ and $\mathrm{C2}$ are the optimal conditions of the queues of the buffers, i.e., the queues  must be at the edge of non-absorption, and constraints $\mathrm{C3}$ and $\mathrm{C4}$ impose the necessary restrictions on the mode selection variables. 

\subsection{Optimal Mode Selection Policy}

In this subsection, we propose the optimal mode selection protocol as a solution to the optimization problem in (\ref{Prob}). In particular, as established in Section II-C, the instantaneous SNRs of the links, $\boldsymbol{\gamma}(i)$, belong to one of the five different SNR regions shown in Fig. \ref{FigOutReg}, i.e., $\mathcal{R}_k,\,\,k=1,\dots,5$. The proposed protocol selects the optimal transmission mode in each time slot based on which SNR region $\boldsymbol{\gamma}(i)$ belongs to. Moreover, the optimal mode selection policy depends on the statistics of the fading gains. Thus, we distinguish several statistical regions for the  fading gains and each statistical region requires a different optimal selection policy.  As we will see later, the optimal mode selection policy  may require rolling a die. Therefore, we define $X_n^M(i)\in \{{1,\dots,M}\}$ as the outcome of rolling the $n$-th die with $M$ faces in the $i$-th time slot. The probabilities of the possible outcomes of rolling the $n$-th die are given by $\Pr\{X_n^M(i)=m\}=p_n^{(m)},\,\, 1\leq m\leq M$.

%%%%%%%%%%%%%%%%%%%%%%%%%%%%%%%%%%%%%%%%%%%%%%%% Table %%%%%%%%%%%%%%%%%%%%%%%%%%%%%%%%%%%%%%%%%
\begin{table*}[!t]
\label{LongVar}
\caption{The Values of the Die Probabilities in Theorem \ref{Prot}.} 
\begin{center}
\scalebox{1}{
\begin{tabular}{|| c | c | c | c | c ||}
  \hline                  
\multirow{2}{*}{ }&\multicolumn{4}{|c||}{$P_{\mathcal{R}_3}\Lequal P_{\mathcal{R}_4}$}  \\ \cline{2-5}
&  $\frac{P_{\mathcal{R}_2}\Minus P_{\mathcal{R}_1}}{P_{\mathcal{R}_3}} \Lequal 0$& $0 \Lequal \frac{P_{\mathcal{R}_2}\Minus P_{\mathcal{R}_1}}{P_{\mathcal{R}_3}} \Lequal 1$ & $1 \Lequal \frac{P_{\mathcal{R}_2}\Minus P_{\mathcal{R}_1}}{P_{\mathcal{R}_3}} \Lequal \frac{2P_{\mathcal{R}_4}}{P_{\mathcal{R}_3}} -1 $ & $\frac{P_{\mathcal{R}_2}\Minus P_{\mathcal{R}_1}}{P_{\mathcal{R}_3}} \Gequal \frac{2P_{\mathcal{R}_4}}{P_{\mathcal{R}_3}} -1$  \\ \hline  

\multirow{1}{*}{Die 1} &$p_1^{(1)}\Equal\frac{1}{2}\Add \frac{P_{\mathcal{R}_2}}{2P_{\mathcal{R}_1}}$&$p_1^{(1)}\Equal1$&$p_1^{(1)}\Equal1$& $p_1^{(1)}\Equal1$ \\ \hline  

\multirow{2}{*}{Die 2} &$p_2^{(1)}\Equal 0$&$p_2^{(1)}\Equal 0$&$p_2^{(1)}\Equal \frac{1}{2}\Minus \frac{P_{\mathcal{R}_1}\Add P_{\mathcal{R}_3}}{2P_{\mathcal{R}_2}}$& $p_2^{(1)}\Equal \frac{1}{3} \Minus \frac{P_{\mathcal{R}_1}\Add 2P_{\mathcal{R}_3}\Minus P_{\mathcal{R}_4}}{3P_{\mathcal{R}_2}}$ \\ 
&$p_2^{(2)}\Equal 0$&$p_2^{(2)}\Equal 0$&$p_2^{(2)}\Equal 0$& $p_2^{(2)}\Equal \frac{1}{3} \Minus \frac{P_{\mathcal{R}_1}\Add 2P_{\mathcal{R}_4}\Minus P_{\mathcal{R}_3}}{3P_{\mathcal{R}_2}}$ \\  \hline

\multirow{2}{*}{Die 3} &$p_3^{(1)}\Equal\frac{P_{\mathcal{R}_4}}{P_{\mathcal{R}_3}}p_4^{(2)}$&$p_3^{(1)}\Equal\frac{P_{\mathcal{R}_4}}{P_{\mathcal{R}_3}}p_4^{(2)} \Add  \frac{P_{\mathcal{R}_2}\Minus P_{\mathcal{R}_1}}{P_{\mathcal{R}_3}}$&$p_3^{(1)}\Equal 1$& $p_3^{(1)}\Equal1$ \\ 
&$p_3^{(2)}\Equal1\Minus p_3^{(1)}$&$p_3^{(2)}\Equal1\Minus p_3^{(1)}$&$p_3^{(2)}\Equal0$&$p_3^{(2)}\Equal0$  \\ \hline

\multirow{2}{*}{Die 4} &\multirow{2}{*}{$p_4^{(1)}\Add p_4^{(2)}\Equal\frac{P_{\mathcal{R}_3}}{P_{\mathcal{R}_4}}$}&\multirow{2}{*}{$p_4^{(1)}\Add p_4^{(2)}\Equal\frac{P_{\mathcal{R}_3}}{P_{\mathcal{R}_4}}$}&$p_4^{(1)}\Equal \frac{P_{\mathcal{R}_2}}{2P_{\mathcal{R}_4}}\Add \frac{P_{\mathcal{R}_3}\Minus P_{\mathcal{R}_1}}{2P_{\mathcal{R}_4}}$& $p_4^{(1)}\Equal 1$ \\ 
&&&$p_4^{(2)}\Equal 0$& $p_4^{(2)}\Equal 0$  \\ \hline
 
% Case 2
                
\multirow{2}{*}{ }&\multicolumn{4}{|c||}{$P_{\mathcal{R}_3}\Gequal P_{\mathcal{R}_4}$}  \\ \cline{2-5} &  $\frac{P_{\mathcal{R}_2}\Minus P_{\mathcal{R}_1}}{P_{\mathcal{R}_4}} \Lequal 0$& $0 \Lequal \frac{P_{\mathcal{R}_2}\Minus P_{\mathcal{R}_1}}{P_{\mathcal{R}_4}} \Lequal 1$ & $1 \Lequal \frac{P_{\mathcal{R}_2}\Minus P_{\mathcal{R}_1}}{P_{\mathcal{R}_4}} \Lequal \frac{2P_{\mathcal{R}_3}}{P_{\mathcal{R}_4}} -1 $ & $\frac{P_{\mathcal{R}_2}\Minus P_{\mathcal{R}_1}}{P_{\mathcal{R}_4}} \Gequal \frac{2P_{\mathcal{R}_3}}{P_{\mathcal{R}_4}} -1$  \\ \hline  

\multirow{1}{*}{Die 1} &$p_1^{(1)}\Equal\frac{1}{2}\Add \frac{P_{\mathcal{R}_2}}{2P_{\mathcal{R}_1}}$&$p_1^{(1)}\Equal1$&$p_1^{(1)}\Equal1$& $p_1^{(1)}\Equal1$ \\ \hline  

\multirow{2}{*}{Die 2} &$p_2^{(1)}\Equal 0$&$p_2^{(1)}\Equal 0$&$p_2^{(1)} \Equal 0 $& $p_2^{(1)}\Equal \frac{1}{3} \Minus \frac{P_{\mathcal{R}_1}\Add 2P_{\mathcal{R}_3}\Minus P_{\mathcal{R}_4}}{3P_{\mathcal{R}_2}}$ \\ 
&$p_2^{(2)}\Equal 0$&$p_2^{(2)}\Equal 0$&$p_2^{(2)}\Equal \frac{1}{2}\Minus \frac{P_{\mathcal{R}_1}\Add P_{\mathcal{R}_4}}{2P_{\mathcal{R}_2}}$& $p_2^{(2)}\Equal \frac{1}{3} \Minus \frac{P_{\mathcal{R}_1}\Add 2P_{\mathcal{R}_4}\Minus P_{\mathcal{R}_3}}{3P_{\mathcal{R}_2}}$ \\  \hline

\multirow{2}{*}{Die 3} &\multirow{2}{*}{$p_3^{(1)}\Add p_3^{(2)}\Equal\frac{P_{\mathcal{R}_4}}{P_{\mathcal{R}_3}}$}&\multirow{2}{*}{$p_3^{(1)}\Add p_3^{(2)}\Equal\frac{P_{\mathcal{R}_4}}{P_{\mathcal{R}_3}}$}&$p_3^{(1)}\Equal \frac{P_{\mathcal{R}_2}}{2P_{\mathcal{R}_3}}\Add \frac{P_{\mathcal{R}_4}\Minus P_{\mathcal{R}_1}}{2P_{\mathcal{R}_3}} $& $p_3^{(1)}\Equal 1$ \\ 
&&&$p_3^{(2)}\Equal 0$& $p_3^{(2)}\Equal 0$  \\ \hline

\multirow{2}{*}{Die 4} &$p_4^{(1)}\Equal\frac{P_{\mathcal{R}_3}}{P_{\mathcal{R}_4}}p_3^{(2)}$&$p_4^{(1)}\Equal\frac{P_{\mathcal{R}_3}}{P_{\mathcal{R}_4}}p_3^{(2)} \Add  \frac{P_{\mathcal{R}_2}\Minus P_{\mathcal{R}_1}}{P_{\mathcal{R}_4}}$&$p_4^{(1)}\Equal 1$& $p_4^{(1)}\Equal1$ \\ 
&$p_4^{(2)}\Equal1\Minus p_4^{(1)}$&$p_4^{(2)}\Equal1\Minus p_4^{(1)}$&$p_4^{(2)}\Equal0$&$p_4^{(2)}\Equal0$  \\ \hline

\end{tabular}
}
\end{center}
\vspace{-0.4cm}
\end{table*}
%%%%%%%%%%%%%%%%%%%%%%%%%%%%%%%%%%%%%%%%%%%%%%%%%%%%%%%%%%%%%%%%%%%%%%%%%%%%%%%%%%%%%%%%%%%%%%%%%%%%%

\begin{theo}\label{Prot}
For $N\to\infty$, the optimal mode selection policy  which maximizes the sum throughput of the considered half-duplex bidirectional relay network with AWGN and block fading is given by
%%%%%%%%%%%%%%%%%%%%%%%%%%%%%%%%%%%%%%%%%%%%%%%% EQUATION %%%%%%%%%%%%%%%%%%%%%%%%%%%%%%%%%%%%%%%%%
\begin{IEEEeqnarray}{lll}
&\text{If} \,\, \boldsymbol{\gamma}(i)\in\mathcal{R}_1 \Longrightarrow \mathcal{M}_{k^*} = 
\begin{cases}
\mathcal{M}_3, \quad &\mathrm{if}\,\,X_1^{2}(i)=1 \\
\mathcal{M}_6, &\mathrm{if}\,\,X_1^{2}(i)=2 
\end{cases} \nonumber \\
&\text{If} \,\, \boldsymbol{\gamma}(i)\in\mathcal{R}_2 \Longrightarrow \mathcal{M}_{k^*} = 
\begin{cases}
\mathcal{M}_1, \quad &\mathrm{if}\,\,X_2^{3}(i)=1 \\
\mathcal{M}_2, &\mathrm{if}\,\,X_2^{3}(i)=2 \\
\mathcal{M}_6, &\mathrm{if}\,\,X_2^{3}(i)=3 
\end{cases} \nonumber \\
&\text{If} \,\, \boldsymbol{\gamma}(i)\in\mathcal{R}_3 \Longrightarrow \mathcal{M}_{k^*} = 
\begin{cases}
\mathcal{M}_1, \quad &\mathrm{if}\,\,X_3^{3}(i)=1 \\
\mathcal{M}_4, &\mathrm{if}\,\,X_3^{3}(i)=2 \\
\mathcal{M}_7, &\mathrm{if}\,\,X_3^{3}(i)=3 
\end{cases} \nonumber \\
&\text{If} \,\, \boldsymbol{\gamma}(i)\in\mathcal{R}_4 \Longrightarrow \mathcal{M}_{k^*} = 
\begin{cases}
\mathcal{M}_2, \quad &\mathrm{if}\,\,X_4^{3}(i)=1 \\
\mathcal{M}_5, &\mathrm{if}\,\,X_4^{3}(i)=2 \\
\mathcal{M}_7, &\mathrm{if}\,\,X_4^{3}(i)=3 
\end{cases} \nonumber \\
&\text{If} \,\, \boldsymbol{\gamma}(i)\in\mathcal{R}_5 \Longrightarrow \mathcal{M}_{k^*} = \mathcal{M}_7
\end{IEEEeqnarray}
%%%%%%%%%%%%%%%%%%%%%%%%%%%%%%%%%%%%%%%%%%%%%%%%%%%%%%%%%%%%%%%%%%%%%%%%%%%%%%%%%%%%%%%%%%%%%%%%%%%%%
where the probabilities for the outcomes of rolling the dice depend on the statistics of the channel gains and are given in Table II.  With this protocol, the maximum sum throughput and minimum system outage probability are given by 
%%%%%%%%%%%%%%%%%%%%%%%%%%%%%%%%%%%%%%%%%%%%%%%% EQUATION %%%%%%%%%%%%%%%%%%%%%%%%%%%%%%%%%%%%%%%%%
\begin{IEEEeqnarray}{lll}
  \bar{R}_{\mathrm{sum}} \hspace{-1mm}
 \Equal \hspace{-1mm}\begin{cases}
 \hspace{-1mm}(P_{\mathcal{R}_1} \Add P_{\mathcal{R}_2} \Add P_{\min})R_0,\hspace{-3mm}&\mathrm{if}\,\frac{P_{\mathcal{R}_2} \Minus P_{\mathcal{R}_1}}{P_{\min}} \Lequal \frac{2P_{\max}}{P_{\min}} \Minus 1\\
 \hspace{-1mm}\frac{2}{3} \hspace{-1mm} \left(\hspace{-0.5mm}2P_{\mathcal{R}_1} \hspace{-1mm} \Add \hspace{-0.5mm} P_{\mathcal{R}_2} \hspace{-1mm} \Add \hspace{-0.5mm} P_{\mathcal{R}_3} \hspace{-1mm} \Add \hspace{-0.5mm} P_{\mathcal{R}_4} \hspace{-0.5mm} \right)\hspace{-1mm} R_0,\hspace{-3mm}&\mathrm{otherwise}
      \end{cases}  \label{MaxThr}\\
      F_{\mathrm{sys}}^{\mathrm{out}} 
 \Equal \begin{cases}
 P_{\mathcal{R}_5} \Add P_{\max},\hspace{-3mm}&\mathrm{if}\,\frac{P_{\mathcal{R}_2} \Minus P_{\mathcal{R}_1}}{P_{\min}} \Lequal \frac{2P_{\max}}{P_{\min}} \Minus 1\\
 \frac{1}{3} - \frac{2}{3}  \left(P_{\mathcal{R}_1} \Minus P_{\mathcal{R}_5}\right),\hspace{-3mm}&\mathrm{otherwise}
      \end{cases} \label{MinFout}
\end{IEEEeqnarray}
%%%%%%%%%%%%%%%%%%%%%%%%%%%%%%%%%%%%%%%%%%%%%%%%%%%%%%%%%%%%%%%%%%%%%%%%%%%%%%%%%%%%%%%%%%%%%%%%%%%%%
where $P_{\max} \Equal \max\{ P_{\mathcal{R}_3}, P_{\mathcal{R}_4}\}$ and $P_{\min} \Equal \min\{ P_{\mathcal{R}_3}, P_{\mathcal{R}_4}\}$. 
\end{theo}

\begin{IEEEproof}
Please refer to Appendix A.
\end{IEEEproof}

We note that the die probabilities in the optimal mode selection policy in Theorem \ref{Prot} depend only on the long term statistics of the
channel gains. Hence, they can be obtained offline and used as long as the channel statistics remain unchanged. Moreover, we give the probabilities of $M-1$ faces for a die with $M$ faces in Table II due to space constraints. The last probability  is obtained as $p_n^{(M)}=1-\sum\nolimits_{m=1}^{M-1}p_n^{(m)}$.

\begin{remk}\label{RemkDiceProb}
We note that for the case $P_{\mathcal{R}_3}\leq P_{\mathcal{R}_4}$ and $\frac{P_{\mathcal{R}_2}-P_{\mathcal{R}_1}}{P_{\mathcal{R}_3}}\leq 1$, we have a degree of freedom in choosing $p_4^{(1)}$ and $p_4^{(2)}$ as only $p_4^{(1)}+p_4^{(2)}=\frac{P_{\mathcal{R}_3}}{P_{\mathcal{R}_4}}$ has to hold, cf. Table II. A similar degree of freedom exists for $p_3^{(1)}$ and $p_3^{(2)}$ for the case $P_{\mathcal{R}_3}\geq P_{\mathcal{R}_4}$ and $\frac{P_{\mathcal{R}_2}-P_{\mathcal{R}_1}}{P_{\mathcal{R}_4}}\leq 1$ as only $p_3^{(1)}+p_3^{(2)}=\frac{P_{\mathcal{R}_4}}{P_{\mathcal{R}_3}}$ has to hold.
\end{remk}

\begin{remk}
We observe that the optimal mode selection policy adopts mode $\mathcal{M}_7$ in instantaneous SNR regions $\mathcal{R}_3$ and $\mathcal{R}_4$ for some channel statistics, cf. Theorem \ref{Prot} and Table II. Hence, although one of the links could support the transmission in this case, the optimal mode selection policy forces all nodes to be silent. The reason for this is that one of the channels is statistically weaker than the other one and the maximum sum throughput is limited by the weaker channel. Therefore, if the transmission modes associated with the stronger channel are always selected, the weaker channel cannot convey the information and consequently constraints $\mathrm{C1}$ and $\mathrm{C2}$ in (\ref{Prob}) are violated.
\end{remk}

\begin{remk}
We assume that the relay is responsible for performing the optimal mode selection using the protocol in Theorem \ref{Prot}. In particular, in the beginning of each time slot, the users send pilots to the relay. The relay has to determine the instantaneous SNR region $\mathcal{R}_l,\,\,l=1,\dots,5$, to select the optimal transmission mode according to Theorem \ref{Prot}. Then, the relay broadcasts the optimal transmission mode to the users using three bits of feedback and  transmission begins. We note that the relay has to determine only the instantaneous SNR region $\mathcal{R}_l$, which is in general a less strict requirement compared to determining the exact values of the instantaneous CSI. Moreover, we assume that even if the relay has the instantaneous CSI, it does not utilize it for adaptive rate transmission due to complexity constraints, e.g., availability of only one fixed coding and modulation scheme and a low feedback overhead requirement. 
\end{remk}

\begin{remk}
Due to data buffering, the proposed protocol introduces an increased end-to-end delay. However, as shown in \cite{NikolaMixed} for one-way relaying, with some modifications of the optimal protocol, the average delay can be bounded  at the expense of  a small loss in the throughput. The delay analysis of the proposed
protocol is beyond the scope of the current paper and is left for future research.
\end{remk}

\subsection{High SNR Analysis}

In the following, we investigate the performance of the proposed protocol in the high SNR regime. To this end, we define $f(x)=o(g(x))$ if $\underset{x\to 0}{\lim} \frac{f(x)}{g(x)} = 0$.

\begin{Corol}\label{CorPout}
The sum throughput and the system outage probability of the protocol in Theorem \ref{Prot} in the high SNR regime, i.e., $\gamma\to\infty$, are given by
\begin{IEEEeqnarray}{lll}
\bar{R}_{\mathrm{sum}} = R_0 \quad \mathrm{and} \quad
      F^{\mathrm{out}}_{\mathrm{sys}} = P_{\max}. 
\end{IEEEeqnarray}
For Rayleigh fading, the outage probability of the system simplifies to
%%%%%%%%%%%%%%%%%%%%%%%%%%%%%%%%%%%%%%%%%%%%%%%% EQUATION %%%%%%%%%%%%%%%%%%%%%%%%%%%%%%%%%%%%%%%%%
\begin{IEEEeqnarray}{ccc}\label{FoutHigh}
       F^{\mathrm{out}}_{\mathrm{sys}}  = \frac{\gamma_{\mathrm{thr}}}{\Omega_{\min}} \cdot \frac{1}{\gamma} + o\left(\frac{1}{\gamma}\right)
\end{IEEEeqnarray}
%%%%%%%%%%%%%%%%%%%%%%%%%%%%%%%%%%%%%%%%%%%%%%%%%%%%%%%%%%%%%%%%%%%%%%%%%%%%%%%%%%%%%%%%%%%%%%%%%%%%%
where $\Omega_{\min}=\min\{\Omega_1,\Omega_2\}$. 
\end{Corol}

\begin{IEEEproof}
For the high SNR regime, i.e., $\gamma\to\infty$, we obtain  $P_{\mathcal{R}_1}\to 1$ and $P_{\mathcal{R}_2}\to 0$ which leads to statistical region $\frac{P_{\mathcal{R}_2} - P_{\mathcal{R}_1}}{P_{\min}} < 0 < \frac{2P_{\max}}{P_{\min}}-1$. Therefore, from (\ref{MaxThr}) and (\ref{MinFout}) and knowing that $\frac{P_{\mathcal{R}_2}+P_{\min}}{P_{\mathcal{R}_1}}\to 0$ and $\frac{P_{\mathcal{R}_5}}{P_{\max}}\to 0$ hold as $\gamma\to \infty$, we obtain $\bar{R}_{\mathrm{sum}} = R_0$ and $F^{\mathrm{out}}_{\mathrm{sys}} = P_{\max}$.

For Rayleigh fading, the probability density functions (pdfs) of $\gamma_1(i)$ and $\gamma_2(i)$ are given by $f_{\gamma_1}(\gamma_1)=\frac{1}{\Omega_1\gamma}e^{-\frac{\gamma_1}{\Omega_1\gamma}}$ and $f_{\gamma_2}(\gamma_2)=\frac{1}{\Omega_2\gamma}e^{-\frac{\gamma_2}{\Omega_2\gamma}}$, respectively. Thus, we obtain $P_{\max}$ as
%%%%%%%%%%%%%%%%%%%%%%%%%%%%%%%%%%%%%%%%%%%%%%%% EQUATION %%%%%%%%%%%%%%%%%%%%%%%%%%%%%%%%%%%%%%%%%
\begin{IEEEeqnarray}{lll}\label{Pmax}
P_{\max} = \left(1-e^{-\frac{\gamma_{\mathrm{thr}}}{\Omega_{\min}\gamma}}\right) e^{-\frac{\gamma_{\mathrm{thr}}}{\Omega_{\max}\gamma}} 
\end{IEEEeqnarray}
%%%%%%%%%%%%%%%%%%%%%%%%%%%%%%%%%%%%%%%%%%%%%%%%%%%%%%%%%%%%%%%%%%%%%%%%%%%%%%%%%%%%%%%%%%%%%%%%%%%%%
where $\Omega_{\min}=\min\{\Omega_1,\Omega_2\}$ and $\Omega_{\max}=\max\{\Omega_1,\Omega_2\}$. Using the Taylor series $e^{x}=1+x+o\left(x\right)$ for $x\to 0$, we obtain $F^{\mathrm{out}}_{\mathrm{sys}}$ in (\ref{FoutHigh}).  This completes the proof.
\end{IEEEproof}

\section{Numerical Results}

In this section, we numerically evaluate the performance of  the  proposed  protocol  for  the  considered  bidirectional
relay network in Rayleigh fading. Moreover, due to space constraints, all presented results are obtained for $\Omega_1=\Omega_2=1$ and $R_0=1$ for all SNR, $\gamma$.

We adopt the traditional two-way, TDBC, and MABC protocols as benchmark schemes \cite{TDBC,MABC}. In order to obtain a fair comparison with respect to the delay, we modify the benchmark schemes such that they also exploit the buffering capability. In this case, although the benchmark
protocols have a fixed and predetermined schedule of transmission, the users are allowed to transmit to the
relay for a fraction of $N$ time slots consecutively and the relay stores the information in its infinite-size buffers. Then, the relay forwards the information to the users in the remaining time slots. The fraction of $N$ time slots allocated to each transmission mode in the benchmark schemes is optimized for maximization of the sum throughput. 

\begin{figure}
\centering
\iftoggle{OneColumn}{%
% OneColumn
\resizebox{0.7\linewidth}{!}{
\psfragfig{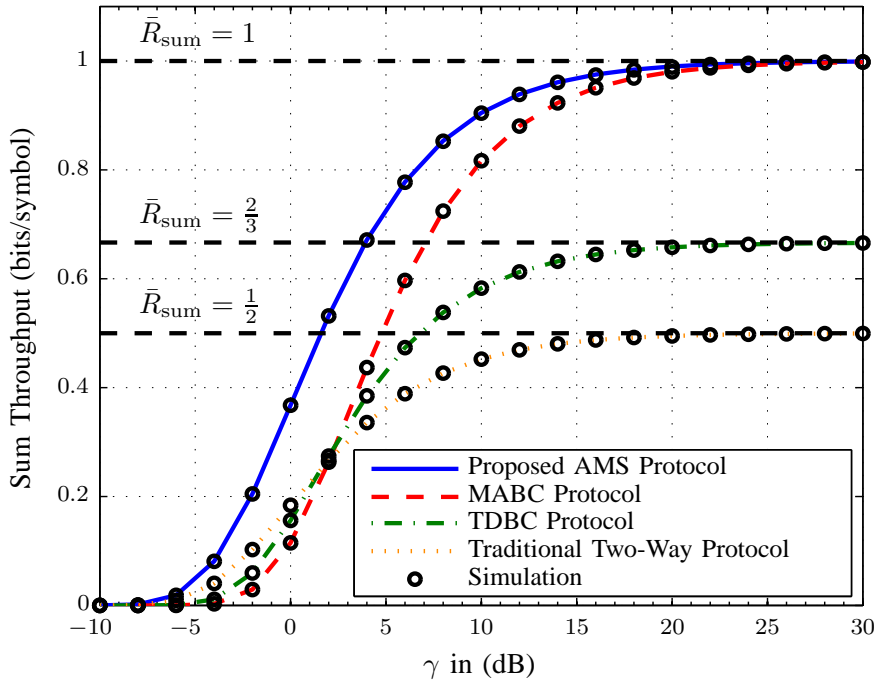}}
}{%
  % TwoColumn
\resizebox{1\linewidth}{!}{
\psfragfig{Fig/SumRate/SumRate}}
}
\vspace{-0.5cm}
\caption{Sum throughput vs. transmit SNR, $\gamma$, in (dB) for $\Omega_1=\Omega_2=1$ and $R_0=1$. Performance comparison between the proposed protocol and benchmark schemes.}
\label{FigThr}
%\vspace{-0.4cm}
\end{figure}

In Fig. \ref{FigThr}, we illustrate the maximum achievable sum throughput, $\bar{R}_{\mathrm{sum}}$, versus the transmit SNR of the nodes, $\gamma$. We observe that the maximum achievable sum throughput is saturated for all protocols at high SNR, i.e., the multiplexing gain is zero because transmit rate $R_0$ is fixed for all SNR. However, the proposed protocol and the MABC protocol achieve a maximum sum throughput of $R_0$ while the TDBC protocol and the traditional two-way protocol achieve  maximum sum throughputs of $\frac{2R_0}{3}$ and $\frac{R_0}{2}$, respectively. Moreover,  the proposed protocol has a superior performance compared to the benchmark schemes for all SNR. 

%%%%%%%%%%%%%%%%%%%%%%%%%%%%%%%%%%%%%%%%%%%%%%%% Figure %%%%%%%%%%%%%%%%%%%%%%%%%%%%%%%%%%%%%%%%%
%%%%%%%%%%%%%%%%%%%%%%%%%%%%%%%%%%%%%%%%%%%%%%%%%%%%%%%%%%%%%%%%%%%%%%%%%%%%%%%%%%%%%%%%%%%%%%%%%

\begin{figure}
\centering
\iftoggle{OneColumn}{%
% OneColumn
\resizebox{0.7\linewidth}{!}{
\psfragfig{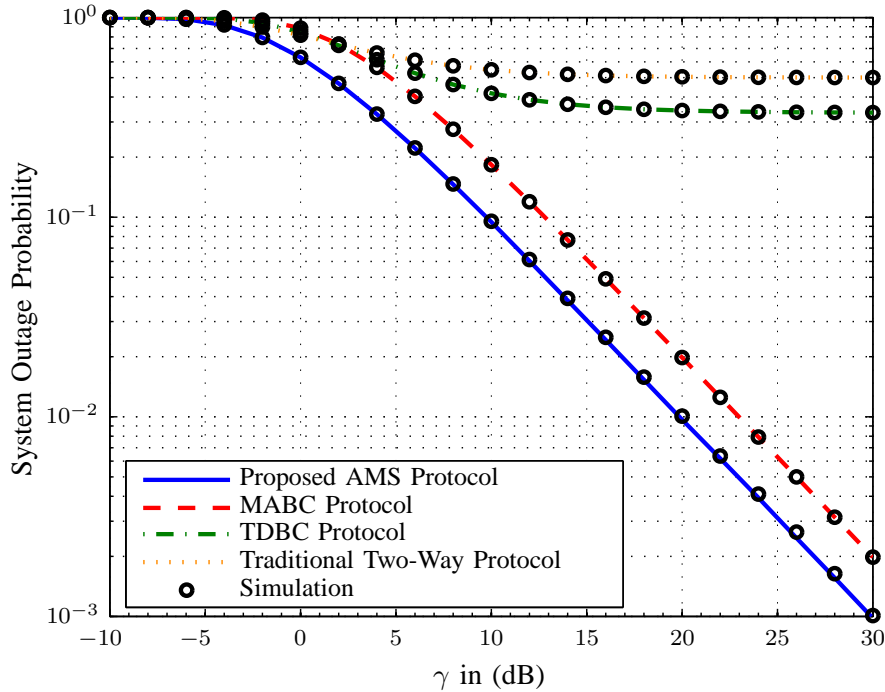}}
}{%
  % TwoColumn
\resizebox{1\linewidth}{!}{
\psfragfig{Fig/SysOutage/SysOutage}}
}
\vspace{-0.5cm}
\caption{System outage probability vs. transmit SNR, $\gamma$, in (dB)  for $\Omega_1=\Omega_2=1$ and $R_0=1$. Performance comparison between the proposed protocol and  benchmark schemes.}
\label{FigOutageSym}
%\vspace{-0.4cm}
\end{figure}

In Fig. \ref{FigOutageSym}, we depict the outage probability of the system versus the transmit SNR of the nodes. We note that the protocol in Theorem \ref{Prot} was derived such that $\bar{R}_{\mathrm{sum}}$ is maximized or equivalently $F^{\mathrm{out}}_{\mathrm{sum}}$ is minimized. However, as stated in Remark \ref{RemkDiceProb}, the per-user throughputs and outage probabilities are not fixed due to the available degree of freedom in choosing the die probabilities.  In Fig. \ref{FigOutageSym}, we also show the per-user outage probabilities $F^{\mathrm{out}}_{12}$ and $F^{\mathrm{out}}_{21}$ of the proposed protocol for the case where the die probabilities are chosen such that rate $\bar{R}_{12}$ has its maximum value. We observe from Fig. \ref{FigOutageSym} that the proposed protocol outperforms the benchmark schemes significantly. In particular, we obtain around $4$ dB SNR gain compared to the best benchmark scheme, the MABC protocol. We note that the per-user outage probabilities can change between the curves of $F^{\mathrm{out}}_{12}$ and $F^{\mathrm{out}}_{21}$ depicted in Fig. \ref{FigOutageSym} depending on how the die probabilities are chosen. 

 %%%%%%%%%%%%%%%%%%%%%%%%%%%%%%%%%%%%%%%%%%%%%%%% Figure %%%%%%%%%%%%%%%%%%%%%%%%%%%%%%%%%%%%%%%%%
%%%%%%%%%%%%%%%%%%%%%%%%%%%%%%%%%%%%%%%%%%%%%%%%%%%%%%%%%%%%%%%%%%%%%%%%%%%%%%%%%%%%%%%%%%%%%%%%%
  
\section{Conclusion}

We derived a protocol which maximizes the sum throughput of bidirectional relay networks with block fading when all nodes transmit with a fixed rate.  The proposed protocol selects the optimal transmission mode in each time slot based on CSIR knowledge. For this to be possible, the relay has to be equipped with two buffers for storage of the information received from the users. We also obtain the diversity-multiplexing trade-off of the proposed protocol. Our numerical results showed that the proposed protocol outperforms the existing protocols significantly in terms of achievable sum throughput and system outage probability.

\iftoggle{OneColumn}{%
% OneColumn

%%%%%%%%%%%%%%%%%%%%%%%%%%%%%%%%%%%%%%%%% Begining One Column Proof %%%%%%%%%%%%%%%%%%%%%%%%%%%%%%%%%%%%%%%%

\appendices

%%%%%%%%%%%%%%%%%%%%%%%%%%%%%%%%%%%%%%%% Appendix B %%%%%%%%%%%%%%%%%%%%%%%%%%%%%%%%%%%%%%%%%%%%%%%%%%%
\section{Proof of Theorem \ref{Prot}}
\label{AppKKT}

In this appendix, we solve the optimization problem given in (\ref{Prob}). We note that because of the binary variables $q_k(i)\in\{0,1\},\,\,\forall i,k$, problem (\ref{Prob}) is an integer program which belongs to the class of non-deterministic polynomial-time hard (NP hard) problems. In this paper, we relax the binary constraint to $0\leq q_k(i)\leq 1$,  which in general implies that the solution of the relaxed problem might not be obtainable with the original problem, i.e., we have a larger feasibility set in the relaxed problem. However, since the relaxed optimization problem is a linear programming problem, an optimal solution is achieved by binary $q_k(i)$, and therefore, the binary relaxation does not affect the maximum achievable sum throughput.  In the following, we investigate the Karush-Kuhn-Tucker (KKT) conditions \cite{Boyd} for the relaxed optimization problem. We note that the relaxed problem is a linear program in the optimization variables $q_k(i)$. Therefore, the KKT conditions are both necessary and sufficient conditions for optimality. To simplify the usage of the KKT conditions, we change the maximization of $\bar{R}_{12}+\bar{R}_{21}$ to the
minimization of $-(\bar{R}_{1r}+\bar{R}_{2r})$ since due to constraints $\mathrm{C1}$ and $\mathrm{C2}$ in (\ref{Prob}), $\bar{R}_{12}=\bar{R}_{r2}=\bar{R}_{1r}$ and $\bar{R}_{21}=\bar{R}_{r1}=\bar{R}_{2r}$ must hold. Moreover, we rewrite all inequality and equality constraints in the form $f(x)\leq 0$ and $g(x)=0$, respectively. Mathematically, we formulate the relaxed optimization problem as
%%%%%%%%%%%%%%%%%%%%%%%%%%%%%%%%%%%%%%%%%%%%%%%% EQUATION %%%%%%%%%%%%%%%%%%%%%%%%%%%%%%%%%%%%%%%%%
\begin{IEEEeqnarray}{Cll}\label{ProbRelaxed}
    {\underset{q_k(i)\,\, \forall i,k}{\mathrm{minimize}}}\,\, & -(\bar{R}_{1r} + \bar{R}_{2r}) \nonumber \\
    \mathrm{subject\,\, to} \,\, &\mathrm{C1}:\quad \bar{R}_{1r}-\bar{R}_{r2} =0 \nonumber \\
    &\mathrm{C2}:\quad \bar{R}_{2r}-\bar{R}_{r1}=0 \nonumber \\
		&\mathrm{C3}:\quad \sum\limits_{k = 1}^7 {q_k}\left( i \right) - 1 = 0, \,\, \forall i   \nonumber \\
    &\mathrm{C4}:\quad - q_k(i) \leq 0, \,\, \forall i, k \nonumber \\
    &\mathrm{C5}:\quad q_k(i) -1 \leq 0, \,\, \forall i, k 
\end{IEEEeqnarray}
%%%%%%%%%%%%%%%%%%%%%%%%%%%%%%%%%%%%%%%%%%%%%%%%%%%%%%%%%%%%%%%%%%%%%%%%%%%%%%%%%%%%%%%%%%%%%%%%%%%%%
The Lagrangian function for the optimization problem in (\ref{ProbRelaxed}) is given by
%%%%%%%%%%%%%%%%%%%%%%%%%%%%%%%%%%%%%%%%%%%%%%%% EQUATION %%%%%%%%%%%%%%%%%%%%%%%%%%%%%%%%%%%%%%%%%
\begin{IEEEeqnarray}{lll}\label{KKTFunction}
   \underset{\mathrm{for}\,\, \forall i,k}{\mathcal{L}(q_k(i),\mu_1, \mu_2,\lambda(i),\alpha_k(i),\beta_k(i))}  = 
 -(\bar{R}_{1r}+\bar{R}_{2r}) + \mu_1(\bar{R}_{1r}-\bar{R}_{r2}) + \mu_2(\bar{R}_{2r}-\bar{R}_{r1})  \nonumber \\
   \qquad \qquad + \mathop \sum \limits_{i = 1}^N \lambda \left( i \right)\left( {\mathop \sum \limits_{k = 1}^7 {q_k}\left( i \right) - 1} \right)
+ \mathop \sum \limits_{i = 1}^N \mathop \sum \limits_{k = 1}^7 {\alpha _k}\left( i \right)\left( {{q_k}\left( i \right) - 1} \right)    - \mathop \sum \limits_{i = 1}^N \mathop \sum \limits_{k = 1}^7 {\beta _k}\left( i \right){q_k}\left( i \right)  \IEEEeqnarraynumspace \IEEEyesnumber
\end{IEEEeqnarray}
%%%%%%%%%%%%%%%%%%%%%%%%%%%%%%%%%%%%%%%%%%%%%%%%%%%%%%%%%%%%%%%%%%%%%%%%%%%%%%%%%%%%%%%%%%%%%%%%%%%%%
where $\mu_1,\mu_2,\lambda(i),\alpha_k(i),$ and $\beta_k(i)$ are the Lagrange multipliers corresponding to constraints $\mathrm{C1,C2,C3,C4}$, and $\mathrm{C5}$, respectively.  

In order to determine the optimal selection policy, $q_k^*(i)$, we must calculate the derivatives of the Lagrangian function in (\ref{KKTFunction}) with respect to $q_k(i)$. This leads to
%%%%%%%%%%%%%%%%%%%%%%%%%%%%%%%%%%%%%%%%%%%%%%%% EQUATION %%%%%%%%%%%%%%%%%%%%%%%%%%%%%%%%%%%%%%%%%
\begin{IEEEeqnarray}{lll}\label{Stationary Mode}
    \frac{\partial\mathcal{L}}{\partial q_1(i)} \Equal - \frac{1}{N}(1\Minus\mu_1)O_1(i)R_{0}\Add \lambda(i)\Add \alpha_1(i)\Minus \beta_1(i) \nonumber  \Equal 0\, \IEEEeqnarraynumspace \IEEEyesnumber \IEEEyessubnumber \\
    \frac{\partial\mathcal{L}}{\partial q_2(i)} \Equal -\frac{1}{N}(1\Minus\mu_2)O_2(i)R_{0}\Add\lambda(i)\Add\alpha_2(i)\Minus\beta_2(i) \nonumber  \Equal 0 \IEEEyessubnumber   \\
    \frac{\partial\mathcal{L}}{\partial q_3(i)} \Equal -\frac{1}{N}(2\Minus \hspace{-0.5mm} \mu_1 \hspace{-0.5mm}\Minus \hspace{-0.5mm}\mu_2)O_3(i)R_{0} \hspace{-0.5mm}\Add\hspace{-0.5mm} \lambda(i) \hspace{-0.5mm}\Add \hspace{-0.5mm} \alpha_3(i)\Minus\beta_3(i) \hspace{-0.5mm}\Equal\hspace{-0.5mm} 0 \qquad  \IEEEyessubnumber \\
    \frac{\partial\mathcal{L}}{\partial q_4(i)} \Equal - \frac{1}{N}\mu_2 O_4(i)R_{0}\Add\lambda(i)\Add\alpha_4(i)\Minus \beta_4(i)\Equal 0 \qquad \, \IEEEyessubnumber \\
    \frac{\partial\mathcal{L}}{\partial q_5(i)} \Equal -\frac{1}{N}\mu_1 O_5(i) R_{0}\Add\lambda(i)\Add\alpha_5(i)\Minus\beta_5(i) \Equal 0 \qquad \, \IEEEyessubnumber \\
    \frac{\partial\mathcal{L}}{\partial q_6(i)} \Equal -\frac{1}{N}(\mu_1\Add\mu_2) O_6(i)R_{0}\Add\lambda(i)\nonumber +\alpha_6(i)\Minus\beta_6(i)\Equal 0 \IEEEyessubnumber \\
    \frac{\partial\mathcal{L}}{\partial q_7(i)} \Equal \lambda(i)\nonumber +\alpha_7(i)\Minus\beta_7(i)\Equal 0. \IEEEyessubnumber
\end{IEEEeqnarray}
%%%%%%%%%%%%%%%%%%%%%%%%%%%%%%%%%%%%%%%%%%%%%%%%%%%%%%%%%%%%%%%%%%%%%%%%%%%%%%%%%%%%%%%%%%%%%%%%%%%%%
Without loss of generality, we first obtain the necessary condition for $q_1^*(i)=1$ and then generalize the result 
to $q_k^*(i)=1,\,\,k=2,\dots,7$. If $q_k^*(i)=1$, from constraint $\mathrm{C3}$ in (\ref{ProbRelaxed}), the other
selection variables are zero, i.e., $q_k^*(i)=0,\,\,k=2,...,7$. Furthermore, from the complementary slackness condition,
we obtain that if an inequality is inactive, the respective Lagrange multiplier must be zero, i.e., $\alpha_k(i)=0,\,\,k = 2,...,7$ and $ \beta_1(i)=0$ have to hold. By substituting these values into (\ref{Stationary Mode}), we obtain
%%%%%%%%%%%%%%%%%%%%%%%%%%%%%%%%%%%%%%%%%%%%%%%% EQUATION %%%%%%%%%%%%%%%%%%%%%%%%%%%%%%%%%%%%%%%%%
\begin{IEEEeqnarray}{lll}\label{MetApp}
    \lambda(i)+\alpha_1(i) = (1-\mu_1)O_1(i)R_{0}  \triangleq \Lambda_1(i) \IEEEyesnumber\IEEEyessubnumber  \\
    \lambda(i)-\beta_2(i) =  (1-\mu_2)O_2(i)R_{0} \triangleq \Lambda_2(i)\IEEEyessubnumber  \\
   \lambda(i)-\beta_3(i) =  (2\Minus \mu_1\Minus \mu_2)O_3(i)R_{0} \triangleq  \Lambda_3(i) \quad\,\,\,\, \IEEEyessubnumber \\
   \lambda(i)-\beta_4(i)  = \mu_2 O_4(i)R_{0} \triangleq \Lambda_4(i) \IEEEyessubnumber\\
    \lambda(i)-\beta_5(i) = \mu_1 O_5(i)R_{0} \triangleq \Lambda_5(i) \IEEEyessubnumber\\
    \lambda(i)-\beta_6(i) = (\mu_1\Add\mu_2) O_6(i)R_{0} \triangleq \Lambda_6(i)\qquad\IEEEyessubnumber \\
     \lambda(i)-\beta_7(i) = 0 \triangleq \Lambda_7(i),\qquad\IEEEyessubnumber
\end{IEEEeqnarray}
%%%%%%%%%%%%%%%%%%%%%%%%%%%%%%%%%%%%%%%%%%%%%%%%%%%%%%%%%%%%%%%%%%%%%%%%%%%%%%%%%%%%%%%%%%%%%%%%%%%%%
where $\Lambda_k(i)$ is referred to as selection metric. By subtracting (\ref{MetApp}b)-(\ref{MetApp}g) from (\ref{MetApp}a), we obtain
%%%%%%%%%%%%%%%%%%%%%%%%%%%%%%%%%%%%%%%%%%%%%%%% EQUATION %%%%%%%%%%%%%%%%%%%%%%%%%%%%%%%%%%%%%%%%%
\begin{IEEEeqnarray}{rCl}\label{eq_2_1}
    \Lambda_1(i) - \Lambda_k(i) = \alpha_1(i)+\beta_k(i), \quad k=2,\dots ,7. \IEEEyesnumber
\end{IEEEeqnarray}
%%%%%%%%%%%%%%%%%%%%%%%%%%%%%%%%%%%%%%%%%%%%%%%%%%%%%%%%%%%%%%%%%%%%%%%%%%%%%%%%%%%%%%%%%%%%%%%%%%%%%
Moreover, the dual feasibility conditions  for Lagrange multipliers imply $\alpha_k(i),\beta_k(i)\geq 0$. By inserting $\alpha_k(i),\beta_k(i)\geq 0$ in (\ref{eq_2_1}), we obtain a necessary condition for $q_1^*(i)=1$ as
%%%%%%%%%%%%%%%%%%%%%%%%%%%%%%%%%%%%%%%%%%%%%%%% EQUATION %%%%%%%%%%%%%%%%%%%%%%%%%%%%%%%%%%%%%%%%%
\begin{IEEEeqnarray}{lll}
    \Lambda_1(i) \geq \max \left \{ \Lambda_2(i), \Lambda_3(i), \Lambda_4(i), \Lambda_5(i), \Lambda_6(i),\Lambda_7(i) \right \}. \quad\IEEEyesnumber
\end{IEEEeqnarray}
%%%%%%%%%%%%%%%%%%%%%%%%%%%%%%%%%%%%%%%%%%%%%%%%%%%%%%%%%%%%%%%%%%%%%%%%%%%%%%%%%%%%%%%%%%%%%%%%%%%%%
Repeating the same procedure for $q_k^*(i)=1,\,\,k=2,\dots,7$, we obtain a necessary condition for selecting transmission mode $\mathcal{M}_{k^*}$ in the $i$-th time slot as 
%%%%%%%%%%%%%%%%%%%%%%%%%%%%%%%%%%%%%%%%%%%%%%% EQUATION %%%%%%%%%%%%%%%%%%%%%%%%%%%%%%%%%%%%%%%%%
\begin{IEEEeqnarray}{lll}\label{OptMet}
   \Lambda_{k^*}(i) \geq {\underset{k\neq k^*}{\max}}\{\Lambda_{k}(i)\}. \IEEEyesnumber
\end{IEEEeqnarray}
%%%%%%%%%%%%%%%%%%%%%%%%%%%%%%%%%%%%%%%%%%%%%%%%%%%%%%%%%%%%%%%%%%%%%%%%%%%%%%%%%%%%%%%%%%%%%%%%%%%%
From (\ref{OptMet}), we conclude that the transmission modes with the largest selection metric values are the candidates for optimal mode selection. In particular, in order to obtain the optimal transmission mode from the candidate modes according to (\ref{OptMet}), we need to answer the following questions: 1) what are the optimal values of $\mu_1$ and $\mu_2$, and 2) given $\mu_1$ and $\mu_2$, how should we select the optimal mode from the modes with identical selection metric values. In the following, we refer to $\mu_1$ and $\mu_2$ also as the selection weights. Next, we investigate the possible values of the selection weights and show that the optimal values of the selection weights depend on the channel statistics. Moreover, we show that for the modes with identical selection metric values, different selection policies that satisfy constraints $\mathrm{C1}$ and $\mathrm{C2}$ in (\ref{Prob}) lead to the same sum throughput. Therefore, for simplicity, we adopt a probabilistic approach via rolling dice where the die probabilities are chosen to satisfy $\mathrm{C1}$ and $\mathrm{C2}$ in (\ref{ProbRelaxed}).

In the following, we first obtain the candidates modes for optimal mode selection in each SNR region given $\mu_1$ and $\mu_2$. Then, we find the optimal values of $\mu_1$ and $\mu_2$ based on the channel statistics.

\subsection{Candidates Modes for Given $\mu_1$ and $\mu_2$}

In order to simplify the analysis, we first obtain the intervals that the optimal values of $\mu_1$ and $\mu_2$ belong to. Then, we obtain the candidates for optimal mode selection for any given $\mu_1$ and $\mu_2$ in the obtained interval. To this end, by comparing the values of the selection metrics, we obtain that if $\mu_1+\mu_2>1$ holds, none of the transmission modes from the users to the relay are selected for all time slots which leads to a violation of constraints $\mathrm{C1}$ and $\mathrm{C2}$ in (\ref{ProbRelaxed}). On the other hand, if $\mu_1+2\mu_2<1$ and  $\mu_2+2\mu_1<1$ hold, none of the transmission modes from the relay to user 1 and from the relay to user 2 are selected, respectively, which lead to a violation of constraints $\mathrm{C1}$ and $\mathrm{C2}$ in (\ref{ProbRelaxed}), respectively. Therefore, we obtain that  $\mu_1+\mu_2\leq 1$, $\mu_1+2\mu_2\geq 1$, and  $2\mu_1+\mu_2\geq 1$ have to hold to fulfill constraints $\mathrm{C1}$ and $\mathrm{C2}$ in (\ref{ProbRelaxed}).  Fig. \ref{FigMUReg} illustrates the plain of $(\mu_1,\mu_2)$ where the shaded area specifies the candidate selection weight values which can lead to the optimal solution. Moreover, we specify the three points  A, B, and C in Fig. \ref{FigMUReg}. These points are the intersection of constraints $\mu_1+\mu_2\leq 1$, $\mu_1+2\mu_2\geq 1$, and  $2\mu_1+\mu_2\geq 1$, and later, we show that these points play an important role for the selection policy.
 
 %%%%%%%%%%%%%%%%%%%%%%%%%%%%%%%%%%%%%%%%%%%%%%%% Figure %%%%%%%%%%%%%%%%%%%%%%%%%%%%%%%%%%%%%%%%%
\begin{figure}
\centering
\iftoggle{OneColumn}{%
% OneColumn
\pstool[width=0.4\linewidth]{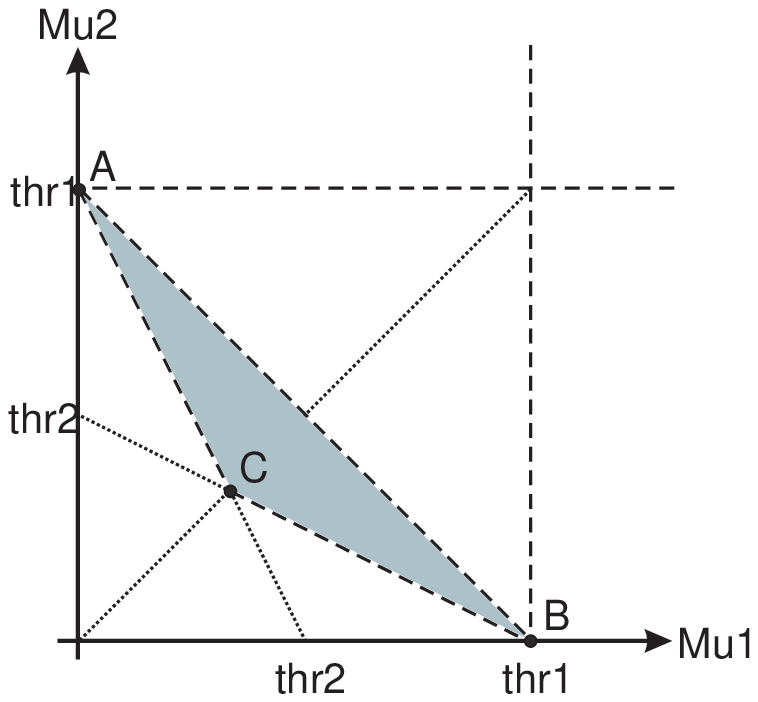}{
\psfrag{Mu1}[c][c][1]{$\mu_1$}
\psfrag{Mu2}[c][c][1]{$\mu_2$}
\psfrag{thr1}[c][c][0.8]{$1$}
\psfrag{thr2}[c][c][0.8]{$\frac{1}{2}$}
\psfrag{A}[c][c][0.8]{\text{A}}
\psfrag{B}[c][c][0.8]{\text{C}}
\psfrag{C}[c][c][0.8]{\text{B}}}
}{%
  % TwoColumn
\pstool[width=0.8\linewidth]{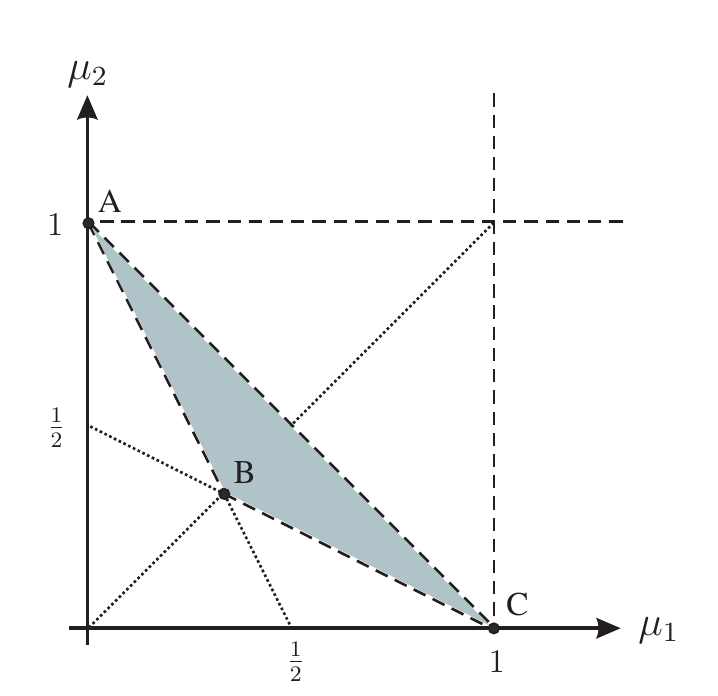}{
\psfrag{Mu1}[c][c][1]{$\mu_1$}
\psfrag{Mu2}[c][c][1]{$\mu_2$}
\psfrag{thr1}[c][c][1]{$1$}
\psfrag{thr2}[c][c][1]{$\frac{1}{2}$}
\psfrag{A}[c][c][1]{\text{A}}
\psfrag{B}[c][c][1]{\text{C}}
\psfrag{C}[c][c][1]{\text{B}}}
}
\caption{Selection weight plain: The shaded area contains the values of $\mu_1$ and $\mu_2$ which are candidates for the optimal values. }
\label{FigMUReg}
\end{figure}
%%%%%%%%%%%%%%%%%%%%%%%%%%%%%%%%%%%%%%%%%%%%%%%%%%%%%%%%%%%%%%%%%%%%%%%%%%%%%%%%%%%%%%%%%%%%%%%%% 

We show how the candidate modes for optimal mode selection are obtained according to (\ref{OptMet}) for SNR region $\mathcal{R}_1$. Then, we extend the result for the other SNR regions $\mathcal{R}_l,\,\,l=2,\dots,5$. For SNR region $\mathcal{R}_1$, we obtained $O_k(i)=1,\,\,k=1,\dots,6$, which leads to the following values for the selection metrics 
%%%%%%%%%%%%%%%%%%%%%%%%%%%%%%%%%%%%%%%%%%%%%%%% EQUATION %%%%%%%%%%%%%%%%%%%%%%%%%%%%%%%%%%%%%%%%%
\begin{IEEEeqnarray}{lll}
    \Lambda_1(i)  = (1-\mu_1)R_{0} \IEEEyesnumber\IEEEyessubnumber  \\
    \Lambda_2(i) =  (1-\mu_2)R_{0}\IEEEyessubnumber  \\
   \Lambda_3(i)  =  \Lambda_1(i)  + \Lambda_2(i) = (2\Minus \mu_1\Minus \mu_2)R_{0}  \IEEEyessubnumber \\
   \Lambda_4(i)  = \mu_2 R_{0}\IEEEyessubnumber\\
   \Lambda_5(i)  = \mu_1  R_{0} \IEEEyessubnumber\\
    \Lambda_6(i) = \Lambda_4(i)  + \Lambda_5(i)=(\mu_1\Add\mu_2)  R_{0} \IEEEyessubnumber \\
    \Lambda_7(i) = 0.\qquad\IEEEyessubnumber
\end{IEEEeqnarray}
%%%%%%%%%%%%%%%%%%%%%%%%%%%%%%%%%%%%%%%%%%%%%%%%%%%%%%%%%%%%%%%%%%%%%%%%%%%%%%%%%%%%%%%%%%%%%%%%%%%%%

Considering that for the optimal $\mu_1$ and $\mu_2$, $\mu_1+\mu_2\leq 1$, $\mu_1+2\mu_2\geq 1$, and  $2\mu_1+\mu_2\geq 1$ hold, we obtain the following candidates for optimal mode selection  for SNR region $\mathcal{R}_1$
%%%%%%%%%%%%%%%%%%%%%%%%%%%%%%%%%%%%%%%%%%%%%%%% EQUATION %%%%%%%%%%%%%%%%%%%%%%%%%%%%%%%%%%%%%%%%%
\begin{IEEEeqnarray}{lllll}\label{SelecCand1}
\text{SNR Region} \,\, \mathcal{R}_1 &\Longrightarrow 
     \begin{cases}
     \mathcal{M}_3, &\mathrm{if}\,\, \mu_1+\mu_2 < 1\\
      \mathcal{M}_3,\mathcal{M}_6, &  \mathrm{if}\,\,\mu_1+\mu_2 = 1, \,\,\mu_1,\mu_2\neq 0,1\\
     \mathcal{M}_1,\mathcal{M}_3,\mathcal{M}_4,\mathcal{M}_6, &\mathrm{if}\,\, \mu_1=0, \,\,\mu_2 = 1 \\
     \mathcal{M}_2,\mathcal{M}_3,\mathcal{M}_5,\mathcal{M}_6, &\mathrm{if}\,\, \mu_1=1, \,\,\mu_2 = 0 
    \end{cases}\qquad\qquad\quad
\end{IEEEeqnarray}
%%%%%%%%%%%%%%%%%%%%%%%%%%%%%%%%%%%%%%%%%%%%%%%%%%%%%%%%%%%%%%%%%%%%%%%%%%%%%%%%%%%%%%%%%%%%%%%%%%%%%
With a similar procedure, we obtain the candidate modes for optimal mode selection for SNR regions $\mathcal{R}_l,\,\,l=2,\dots,5$ as follows
%%%%%%%%%%%%%%%%%%%%%%%%%%%%%%%%%%%%%%%%%%%%%%%% EQUATION %%%%%%%%%%%%%%%%%%%%%%%%%%%%%%%%%%%%%%%%%
\begin{IEEEeqnarray}{lllll}\label{SelecCand2}
\text{SNR Region} \,\, \mathcal{R}_2 &\Longrightarrow 
     \begin{cases}
     \mathcal{M}_6, & \mathrm{if}\,\,\mu_1+2\mu_2 > 1,\,\,2\mu_1+\mu_2 > 1\\
     \mathcal{M}_1,\mathcal{M}_6, & \mathrm{if}\,\,\mu_1+2\mu_2 = 1,\,\,2\mu_1+\mu_2 > 1,\,\,\mu_1\neq 1,\,\,\mu_2\neq 0 \\
      \mathcal{M}_2,\mathcal{M}_6, & \mathrm{if}\,\,\mu_1+2\mu_2 > 1,\,\,2\mu_1+\mu_2 = 1,\,\,\mu_1\neq 0,\,\,\mu_2\neq 1\\
      \mathcal{M}_1,\mathcal{M}_2,\mathcal{M}_6, & \mathrm{if}\,\,\mu_1+2\mu_2 = 1,\,\,2\mu_1+\mu_2 = 1\\
     \mathcal{M}_1,\mathcal{M}_4,\mathcal{M}_6, & \mathrm{if}\,\,\mu_1=0, \,\,\mu_2 = 1 \\
     \mathcal{M}_2,\mathcal{M}_5,\mathcal{M}_6, & \mathrm{if}\,\,\mu_1=1, \,\,\mu_2 = 0 
    \end{cases} \IEEEyesnumber \IEEEyessubnumber  \\
\text{SNR Region} \,\, \mathcal{R}_3 &\Longrightarrow 
     \begin{cases}
     \mathcal{M}_1, & \mathrm{if}\,\, \mu_1 +\mu_2 < 1\\
     \mathcal{M}_1,\mathcal{M}_4, &  \mathrm{if}\,\, \mu_1 +\mu_2 = 1,\,\,\mu_1\neq 1,\,\, \mu_2\neq 0\\
     \mathcal{M}_1, \mathcal{M}_4, \mathcal{M}_7, &  \mathrm{if}\,\, \mu_1=1, \,\,\mu_2 = 0  
    \end{cases} \IEEEyessubnumber  \\
    \text{SNR Region} \,\, \mathcal{R}_4 &\Longrightarrow 
     \begin{cases}
     \mathcal{M}_2, & \mathrm{if}\,\, \mu_1 +\mu_2 < 1\\
     \mathcal{M}_2,\mathcal{M}_5, &  \mathrm{if}\,\, \mu_1 +\mu_2 = 1,\,\,\mu_1\neq 0,\,\, \mu_2\neq 1\\
     \mathcal{M}_2, \mathcal{M}_5, \mathcal{M}_7, &  \mathrm{if}\,\, \mu_1 = 0, \,\,\mu_2 = 1 
    \end{cases} \IEEEyessubnumber  \\
    \text{SNR Region} \,\, \mathcal{R}_5 &\Longrightarrow \mathcal{M}_7  \IEEEyessubnumber 
\end{IEEEeqnarray}
%%%%%%%%%%%%%%%%%%%%%%%%%%%%%%%%%%%%%%%%%%%%%%%%%%%%%%%%%%%%%%%%%%%%%%%%%%%%%%%%%%%%%%%%%%%%%%%%%%%%%

\subsection{Optimal Values of $\mu_1$ and $\mu_2$}

As we see from (\ref{SelecCand1}) and (\ref{SelecCand2}), for the optimal mode selection policy, we are not required to know the the exact values of the selection weights. In particular, we only need to know whether the optimal selection weight pair, $(\mu_1,\mu_2)$, is points A, B, or C, a point on the lines between points A, B, and C, or a point in the interior of the shaded area in Fig. \ref{FigMUReg}. In this subsecction, we investigate the the possible cases for $\mu_1$ and $\mu_2$.

\noindent
\textbf{Point A:} For this point, we have $\mu_1=0$ and $\mu_2=1$, see Fig. \ref{FigMUReg}. Then, according to (\ref{SelecCand1}) and (\ref{SelecCand2}), the optimal mode selection policy may select only 1) modes $\mathcal{M}_1$, $\mathcal{M}_3$, $\mathcal{M}_4$, and $\mathcal{M}_6$ in SNR region $\mathcal{R}_1$, 2) modes $\mathcal{M}_1$, $\mathcal{M}_4$, and $\mathcal{M}_6$ in SNR region $\mathcal{R}_2$, 3) modes $\mathcal{M}_1$ and $\mathcal{M}_4$ in SNR region $\mathcal{R}_3$, 4) modes $\mathcal{M}_2$, $\mathcal{M}_5$, and $\mathcal{M}_7$ in SNR region $\mathcal{R}_4$, and 5) mode $\mathcal{M}_7$ in SNR region $\mathcal{R}_5$.  In order to determine which of the candidate modes  should be selected in each SNR region, we propose a probabilistic approach via rolling a die. This approach leads to the maximum throughput. We note that the optimal solution is not unique and the maximum throughput can be also achieved via other strategies instead of the probabilistic one. However, a probabilistic approach emphasizes that the value of $q_k(i)$ in each time slot is binary and therefore, the binary relaxation does not change the maximum sum throughput.  Moreover, the die probabilities are obtained such that constraints $\mathrm{C1}$ and $\mathrm{C2}$ in (\ref{ProbRelaxed}) hold. Knowing that modes $\mathcal{M}_3$ and $\mathcal{M}_6$ are more spectrum efficient than the point-to-point modes and in order to simplify the derivation of the die probabilities, we consider two cases based on the selection of the point-to-point modes in SNR regions $\mathcal{R}_1$ and $\mathcal{R}_2$. Specifically, if for some channel statistics, we can select modes $\mathcal{M}_3$ and $\mathcal{M}_6$ instead of the point-to-point modes in SNR regions $\mathcal{R}_1$ and $\mathcal{R}_2$, where constraints $\mathrm{C1}$ and $\mathrm{C2}$ in (\ref{ProbRelaxed}) hold, the selection of the point-to-point modes reduces the sum throughput. Therefore, the optimal mode selection selects modes $\mathcal{M}_3$ and $\mathcal{M}_6$ instead of the point-to-point modes unless constraints $\mathrm{C1}$ and $\mathrm{C2}$ in (\ref{ProbRelaxed}) cannot be fulfilled.  In the following, we consider the cases where the point-to-point modes are not selected (Case 1) and are selected (Case 2) in SNR regions $\mathcal{R}_1$ and $\mathcal{R}_2$.  derive the die probabilities and the corresponding conditions on the channel statistics.

\textbf{Case 1:} In this case, the optimal selection policy involves only modes $\mathcal{M}_3$ and $\mathcal{M}_6$ in SNR regions $\mathcal{R}_1$ and $\mathcal{R}_2$.  In particular, considering the probabilistic approach and according to (\ref{SelecCand1}) and (\ref{SelecCand2}), the optimal mode selection policy selects 1) modes $\mathcal{M}_3$ and $\mathcal{M}_6$ with probabilities $p_{36}$ and $1-p_{36}$, respectively, in SNR region $\mathcal{R}_1$,  2) mode $\mathcal{M}_6$ in SNR region $\mathcal{R}_2$, 3) modes $\mathcal{M}_1$ and $\mathcal{M}_4$ with probabilities $p_{14}$  and $1-p_{14}$, respectively, in SNR region $\mathcal{R}_3$,  4) modes $\mathcal{M}_2$, $\mathcal{M}_5$, and $\mathcal{M}_7$ with probabilities  $p_{257}$, $p_{527}$, and $1-p_{257}-p_{527}$, respectively, in SNR region $\mathcal{R}_4$, and 5) mode $\mathcal{M}_7$  in SNR region $\mathcal{R}_7$. Then, constraints $\mathrm{C1}$ and $\mathrm{C2}$ in (\ref{ProbRelaxed})  lead to the following equations
%%%%%%%%%%%%%%%%%%%%%%%%%%%%%%%%%%%%%%%%%%%%%%%% EQUATION %%%%%%%%%%%%%%%%%%%%%%%%%%%%%%%%%%%%%%%%%
\begin{IEEEeqnarray}{lll}\label{ConMU1-2}
     P_{\mathcal{R}_1}p_{36} +P_{\mathcal{R}_3} p_{14} = P_{\mathcal{R}_1}(1-p_{36}) + P_{\mathcal{R}_2} + P_{\mathcal{R}_4} p_{527}  \IEEEyesnumber\IEEEyessubnumber\\
   P_{\mathcal{R}_1}p_{36} + P_{\mathcal{R}_4}p_{257} = P_{\mathcal{R}_1}(1-p_{36}) + P_{\mathcal{R}_2} + P_{\mathcal{R}_3}(1-p_{14}), \IEEEyessubnumber 
\end{IEEEeqnarray}
%%%%%%%%%%%%%%%%%%%%%%%%%%%%%%%%%%%%%%%%%%%%%%%%%%%%%%%%%%%%%%%%%%%%%%%%%%%%%%%%%%%%%%%%%%%%%%%%%%%%%
In order to satisfy the conditions in (\ref{ConMU1-2}), we obtain that $P_{\mathcal{R}_3} \leq P_{\mathcal{R}_4}$ and $\frac{P_{\mathcal{R}_2}-P_{\mathcal{R}_1}}{P_{\mathcal{R}_3}} \leq 1$ must hold. Moreover, the  die probabilities are given by
%%%%%%%%%%%%%%%%%%%%%%%%%%%%%%%%%%%%%%%%%%%%%%%% EQUATION %%%%%%%%%%%%%%%%%%%%%%%%%%%%%%%%%%%%%%%%%
\begin{IEEEeqnarray}{lcl} \label{ProbP3LessP4}
\mathrm{if}\,\, &\frac{P_{\mathcal{R}_2}-P_{\mathcal{R}_1}}{P_{\mathcal{R}_3}} \leq 0 &\Longrightarrow 
     \begin{cases}
     p_{36} = \frac{1}{2}+\frac{P_{\mathcal{R}_2}}{2P_{\mathcal{R}_1}}\\
     p_{14}=\frac{P_{\mathcal{R}_4}}{P_{\mathcal{R}_3}}p_{527}\\
     p_{257}+p_{527}=\frac{P_{\mathcal{R}_3}}{P_{\mathcal{R}_4}}
    \end{cases}  \IEEEyesnumber\IEEEyessubnumber\\
    \mathrm{if}\,\, &0\leq \frac{P_{\mathcal{R}_2}-P_{\mathcal{R}_1}}{P_{\mathcal{R}_3}} \leq 1 &\Longrightarrow
     \begin{cases}
     p_{36} =  1\\
     p_{14}=\frac{P_{\mathcal{R}_4}}{P_{\mathcal{R}_3}}p_{527}+\frac{P_{\mathcal{R}_2}-P_{\mathcal{R}_1}}{P_{\mathcal{R}_3}}\\
     p_{257}+p_{527}=\frac{P_{\mathcal{R}_3}}{P_{\mathcal{R}_4}}
    \end{cases} \IEEEyessubnumber
\end{IEEEeqnarray}
%%%%%%%%%%%%%%%%%%%%%%%%%%%%%%%%%%%%%%%%%%%%%%%%%%%%%%%%%%%%%%%%%%%%%%%%%%%%%%%%%%%%%%%%%%%%%%%%%%%%%
The maximum sum throughput and the minimum system outage probability  
are given by
%%%%%%%%%%%%%%%%%%%%%%%%%%%%%%%%%%%%%%%%%%%%%%%% EQUATION %%%%%%%%%%%%%%%%%%%%%%%%%%%%%%%%%%%%%%%%%
\begin{IEEEeqnarray}{ccc}
 \bar{R}_{\mathrm{sum}} =
      \left(P_{\mathcal{R}_1} + P_{\mathcal{R}_2} + P_{\mathcal{R}_3} \right)R_0  \label{Thr1} \\
 F_{\mathrm{sys}}^{\mathrm{out}} =
       P_{\mathcal{R}_4} + P_{\mathcal{R}_5},\label{Fout1}
\end{IEEEeqnarray}
%%%%%%%%%%%%%%%%%%%%%%%%%%%%%%%%%%%%%%%%%%%%%%%%%%%%%%%%%%%%%%%%%%%%%%%%%%%%%%%%%%%%%%%%%%%%%%%%%%%%%
respectively.

%Moreover, the respective outage probabilities for both transmission directions are given by
%%%%%%%%%%%%%%%%%%%%%%%%%%%%%%%%%%%%%%%%%%%%%%%%% EQUATION %%%%%%%%%%%%%%%%%%%%%%%%%%%%%%%%%%%%%%%%%
%\begin{IEEEeqnarray}{ccc}\label{OutHigh}
%      F^{\mathrm{out}}_{12} = \begin{cases}
%      1-P_{\mathcal{R}_1} - P_{\mathcal{R}_2} -2p_{527}  P_{\mathcal{R}_4}, &\mathrm{if}\,\, \frac{P_{\mathcal{R}_2}-P_{\mathcal{R}_1}}{P_{\mathcal{R}_3}} \leq 0\\
%      1-2P_{\mathcal{R}_2} - 2p_{527}  P_{\mathcal{R}_4}, &\mathrm{if}\,\, 0\leq \frac{P_{\mathcal{R}_2}-P_{\mathcal{R}_1}}{P_{\mathcal{R}_3}} \leq 1
%      \end{cases}  \IEEEyesnumber\IEEEyessubnumber \\
%        F^{\mathrm{out}}_{21} = \begin{cases}
%      1-P_{\mathcal{R}_1} - P_{\mathcal{R}_2} -2p_{257}  P_{\mathcal{R}_4}, &\mathrm{if}\,\, \frac{P_{\mathcal{R}_2}-P_{\mathcal{R}_1}}{P_{\mathcal{R}_3}} \leq 0\\
%      1-2P_{\mathcal{R}_2} - 2p_{257}  P_{\mathcal{R}_4}, &\mathrm{if}\,\, 0\leq \frac{P_{\mathcal{R}_2}-P_{\mathcal{R}_1}}{P_{\mathcal{R}_3}} \leq 1
%      \end{cases}.  \IEEEyessubnumber  
%\end{IEEEeqnarray}
%%%%%%%%%%%%%%%%%%%%%%%%%%%%%%%%%%%%%%%%%%%%%%%%%%%%%%%%%%%%%%%%%%%%%%%%%%%%%%%%%%%%%%%%%%%%%%%%%%%%%%

\textbf{Case 2:} From (\ref{ProbP3LessP4}b), if $\frac{P_{\mathcal{R}_2}-P_{\mathcal{R}_1}}{P_{\mathcal{R}_3}} = 1$ holds, we obtain $p_{36}=1,p_{14}=1$, and $p_{527}=0$. Intuitively, we can conclude that for $\frac{P_{\mathcal{R}_2}-P_{\mathcal{R}_1}}{P_{\mathcal{R}_3}} > 1$, the transmission from user 1 to the relay in regions $\mathcal{R}_1$ and $\mathcal{R}_3$ is not enough to satisfy constraint $\mathrm{C1}$ in (\ref{ProbRelaxed}). Therefore, the optimal selection policy requires to select point-to-point mode $\mathcal{M}_1$ in SNR region $\mathcal{R}_2$. Hence, for channel statistics satisfying $\frac{P_{\mathcal{R}_2}-P_{\mathcal{R}_1}}{P_{\mathcal{R}_3}} > 1$, the optimal mode selection policy selects 1) mode $\mathcal{M}_3$ in SNR region $\mathcal{R}_1$,  2) modes $\mathcal{M}_1$ and  $\mathcal{M}_6$ with probabilities $1-p_{16}$ and $p_{16}$, respectively, in SNR region $\mathcal{R}_2$,  3) mode $\mathcal{M}_1$ in SNR region $\mathcal{R}_3$,  4) modes $\mathcal{M}_2$ and $\mathcal{M}_7$ with probability $p_{27}$ and $1-p_{27}$, respectively, in SNR region $\mathcal{R}_4$, and 5) mode $\mathcal{M}_7$ in SNR region $\mathcal{R}_5$. In particular, constraints $\mathrm{C1}$ and $\mathrm{C2}$ in (\ref{ProbRelaxed})  lead to the following equations
%%%%%%%%%%%%%%%%%%%%%%%%%%%%%%%%%%%%%%%%%%%%%%%% EQUATION %%%%%%%%%%%%%%%%%%%%%%%%%%%%%%%%%%%%%%%%%
\begin{IEEEeqnarray}{lll}   
    P_{\mathcal{R}_1} + P_{\mathcal{R}_2} p_{16} + P_{\mathcal{R}_3} = P_{\mathcal{R}_2}(1-p_{16})  \IEEEyesnumber\IEEEyessubnumber\\    
    P_{\mathcal{R}_1}  + P_{\mathcal{R}_4}p_{27} = P_{\mathcal{R}_2}(1-p_{16}), \IEEEyessubnumber 
\end{IEEEeqnarray}
%%%%%%%%%%%%%%%%%%%%%%%%%%%%%%%%%%%%%%%%%%%%%%%%%%%%%%%%%%%%%%%%%%%%%%%%%%%%%%%%%%%%%%%%%%%%%%%%%%%%%
In order to satisfy the above equations,   $P_{\mathcal{R}_3} \leq P_{\mathcal{R}_4}$ and $1\leq \frac{P_{\mathcal{R}_2}-P_{\mathcal{R}_1}}{P_{\mathcal{R}_3}} \leq \frac{2P_{\mathcal{R}_4}}{P_{\mathcal{R}_3}}-1$ must hold. Moreover, the  die probabilities are obtained as 
%%%%%%%%%%%%%%%%%%%%%%%%%%%%%%%%%%%%%%%%%%%%%%%% EQUATION %%%%%%%%%%%%%%%%%%%%%%%%%%%%%%%%%%%%%%%%%
\begin{IEEEeqnarray}{lcl}
     \begin{cases}
     p_{16}=\frac{1}{2}-\frac{P_{\mathcal{R}_1}+P_{\mathcal{R}_3}}{2P_{\mathcal{R}_2}}\\
     p_{27}=\frac{P_{\mathcal{R}_2}}{2P_{\mathcal{R}_4}}+\frac{P_{\mathcal{R}_3}-P_{\mathcal{R}_1}}{2P_{\mathcal{R}_4}}
    \end{cases}  
\end{IEEEeqnarray}
%%%%%%%%%%%%%%%%%%%%%%%%%%%%%%%%%%%%%%%%%%%%%%%%%%%%%%%%%%%%%%%%%%%%%%%%%%%%%%%%%%%%%%%%%%%%%%%%%%%%%
       
The maximum sum throughput and the minimum  system outage probability  are given by
%%%%%%%%%%%%%%%%%%%%%%%%%%%%%%%%%%%%%%%%%%%%%%%% EQUATION %%%%%%%%%%%%%%%%%%%%%%%%%%%%%%%%%%%%%%%%%
\begin{IEEEeqnarray}{ccc}
 \bar{R}_{\mathrm{sum}} =
      \left(P_{\mathcal{R}_1} + P_{\mathcal{R}_2} + P_{\mathcal{R}_3} \right)R_0   \label{Thr2}\\
      F^{\mathrm{out}}_{\mathrm{sys}} = P_{\mathcal{R}_4} + P_{\mathcal{R}_5},  \label{Fout2}
\end{IEEEeqnarray}
%%%%%%%%%%%%%%%%%%%%%%%%%%%%%%%%%%%%%%%%%%%%%%%%%%%%%%%%%%%%%%%%%%%%%%%%%%%%%%%%%%%%%%%%%%%%%%%%%%%%%
respectively.

\noindent
\textbf{Point B:} For this point, we have $\mu_1=\mu_2=\frac{1}{3}$. Then, according to (\ref{SelecCand1}) and (\ref{SelecCand2}), the optimal mode selection policy selects only 1) mode $\mathcal{M}_3$ in SNR region $\mathcal{R}_1$, 2) modes $\mathcal{M}_1$, $\mathcal{M}_2$, and $\mathcal{M}_6$ with probabilities $p_{126}$, $p_{216}$, and $1-p_{126}-p_{216}$, respectively, in SNR region $\mathcal{R}_2$, 3) mode $\mathcal{M}_1$ in SNR region $\mathcal{R}_3$, 4) mode $\mathcal{M}_2$ in SNR region $\mathcal{R}_4$, and 5) mode $\mathcal{M}_7$ in SNR region $\mathcal{R}_5$. Then, constraints $\mathrm{C1}$ and $\mathrm{C2}$ in (\ref{ProbRelaxed})  lead to the following equations
%%%%%%%%%%%%%%%%%%%%%%%%%%%%%%%%%%%%%%%%%%%%%%%% EQUATION %%%%%%%%%%%%%%%%%%%%%%%%%%%%%%%%%%%%%%%%%
\begin{IEEEeqnarray}{lll}     
    P_{\mathcal{R}_1} + P_{\mathcal{R}_2} p_{126} + P_{\mathcal{R}_3} = P_{\mathcal{R}_2}(1-p_{126}-p_{216})   \IEEEyesnumber\IEEEyessubnumber\\     
    P_{\mathcal{R}_1} + P_{\mathcal{R}_2} p_{216} + P_{\mathcal{R}_4} = P_{\mathcal{R}_2}(1-p_{126}-p_{216}),   \IEEEyessubnumber
\end{IEEEeqnarray}
%%%%%%%%%%%%%%%%%%%%%%%%%%%%%%%%%%%%%%%%%%%%%%%%%%%%%%%%%%%%%%%%%%%%%%%%%%%%%%%%%%%%%%%%%%%%%%%%%%%%%
In order to satisfy the above equations, $\frac{P_{\mathcal{R}_2}-P_{\mathcal{R}_1}}{P_{\mathcal{R}_3}} \geq \frac{2P_{\mathcal{R}_4}}{P_{\mathcal{R}_3}}-1$  must hold for $P_{\mathcal{R}_3} \leq P_{\mathcal{R}_4}$, and $\frac{P_{\mathcal{R}_2}-P_{\mathcal{R}_1}}{P_{\mathcal{R}_4}} \geq \frac{2P_{\mathcal{R}_3}}{P_{\mathcal{R}_4}}-1$  must hold for $P_{\mathcal{R}_3} \geq P_{\mathcal{R}_4}$. Moreover, the die probabilities are obtained as
%%%%%%%%%%%%%%%%%%%%%%%%%%%%%%%%%%%%%%%%%%%%%%%% EQUATION %%%%%%%%%%%%%%%%%%%%%%%%%%%%%%%%%%%%%%%%%
\begin{IEEEeqnarray}{lll}
    p_{126} =  \frac{1}{3} -\frac{P_{\mathcal{R}_1}+2P_{\mathcal{R}_3}-P_{\mathcal{R}_4}}{3P_{\mathcal{R}_2}}  \IEEEyesnumber\IEEEyessubnumber\\
    p_{216} =  \frac{1}{3} -\frac{P_{\mathcal{R}_1}+2P_{\mathcal{R}_4}-P_{\mathcal{R}_3}}{3P_{\mathcal{R}_2}}   \IEEEyessubnumber
\end{IEEEeqnarray}
%%%%%%%%%%%%%%%%%%%%%%%%%%%%%%%%%%%%%%%%%%%%%%%%%%%%%%%%%%%%%%%%%%%%%%%%%%%%%%%%%%%%%%%%%%%%%%%%%%%%%

The maximum sum throughput and the minimum system outage probability are given by
%%%%%%%%%%%%%%%%%%%%%%%%%%%%%%%%%%%%%%%%%%%%%%%% EQUATION %%%%%%%%%%%%%%%%%%%%%%%%%%%%%%%%%%%%%%%%%
\begin{IEEEeqnarray}{ccc}
 \bar{R}_{\mathrm{sum}}  = \frac{2}{3}
      \left(2P_{\mathcal{R}_1} + P_{\mathcal{R}_2} + P_{\mathcal{R}_3} + P_{\mathcal{R}_4}\right)R_0 \label{Thr3} \\
      F_{\mathrm{sys}}^{\mathrm{out}}  = \frac{1}{3} - \frac{2}{3}  \left(P_{\mathcal{R}_1} \Minus P_{\mathcal{R}_5}\right), \label{Fout3}
\end{IEEEeqnarray}
%%%%%%%%%%%%%%%%%%%%%%%%%%%%%%%%%%%%%%%%%%%%%%%%%%%%%%%%%%%%%%%%%%%%%%%%%%%%%%%%%%%%%%%%%%%%%%%%%%%%%
respectively.

%Moreover, the respective outage probabilities for both transmission directions are given by
%%%%%%%%%%%%%%%%%%%%%%%%%%%%%%%%%%%%%%%%%%%%%%%%% EQUATION %%%%%%%%%%%%%%%%%%%%%%%%%%%%%%%%%%%%%%%%%
%\begin{IEEEeqnarray}{ccc}
%      F^{\mathrm{out}}_{12} = \frac{1}{3} - \frac{2}{3} \left( P_{\mathcal{R}_1}  - 3 P_{\mathcal{R}_3} - P_{\mathcal{R}_5}\right) \IEEEyesnumber\IEEEyessubnumber\\
%      F^{\mathrm{out}}_{21} = \frac{1}{3} - \frac{2}{3} \left( P_{\mathcal{R}_1}  - 3 P_{\mathcal{R}_4} - P_{\mathcal{R}_5}\right)  \IEEEyessubnumber
%\end{IEEEeqnarray}
%%%%%%%%%%%%%%%%%%%%%%%%%%%%%%%%%%%%%%%%%%%%%%%%%%%%%%%%%%%%%%%%%%%%%%%%%%%%%%%%%%%%%%%%%%%%%%%%%%%%%%

\noindent
\textbf{Point C:} For this point, we have $\mu_1=1$ and $\mu_2=0$. The optimal selection policy for point C is similar to that for point A, if we change the roles of $\mu_1$ and $\mu_2$, mode $\mathcal{M}_1$ and $\mathcal{M}_1$, mode $\mathcal{M}_4$ and $\mathcal{M}_5$. Due to space constraint, we exclude the detailed derivation of the optimal selection policy for this point.

The analysis of the points on the line A-B is similar to that for point B when $p_{216}=0$. For these points to be optimal, $\frac{P_{\mathcal{R}_2}-P_{\mathcal{R}_1}}{P_{\mathcal{R}_3}} = \frac{2P_{\mathcal{R}_4}}{P_{\mathcal{R}_3}}-1$ and $\frac{P_{\mathcal{R}_2}-P_{\mathcal{R}_1}}{P_{\mathcal{R}_4}} \geq \frac{2P_{\mathcal{R}_3}}{P_{\mathcal{R}_4}}-1$ have to hold. Similarly, the analysis of the points on the line C-B is similar to that for point B when $p_{126}=0$. For these points to be optimal,  $\frac{P_{\mathcal{R}_2}-P_{\mathcal{R}_1}}{P_{\mathcal{R}_4}} = \frac{2P_{\mathcal{R}_3}}{P_{\mathcal{R}_4}}-1$ and $\frac{P_{\mathcal{R}_2}-P_{\mathcal{R}_1}}{P_{\mathcal{R}_3}} \geq \frac{2P_{\mathcal{R}_4}}{P_{\mathcal{R}_3}}-1$ have to hold. In a similar manner, the analysis for the points in the interior of the shaded area in Fig. \ref{FigOutReg}, is similar  to the one for point B when $p_{216}=p_{216}=0$. For these points to be optimal, $\frac{P_{\mathcal{R}_2}-P_{\mathcal{R}_1}}{P_{\mathcal{R}_4}} = \frac{2P_{\mathcal{R}_3}}{P_{\mathcal{R}_4}}-1$ and $\frac{P_{\mathcal{R}_2}-P_{\mathcal{R}_1}}{P_{\mathcal{R}_3}} = \frac{2P_{\mathcal{R}_4}}{P_{\mathcal{R}_3}}-1$ have to hold.

We have now investigated all possible values of the selection weights and the necessary conditions for the optimality of each pair. Moreover, the necessary conditions obtained for the maximum sum throughput in (\ref{Thr1}), (\ref{Thr2}), and (\ref{Thr3}) are mutually exclusive which leads to the conclusion that the obtained conditions are indeed sufficient. These mutually exclusive conditions on the channel statistics are specified in Table II. To summarize and provide a compact solution, we introduce  $X_n^M(i)\in \{{1,\dots,M}\}$ as the outcome of rolling the $n$-th die with $M$ faces in the $i$-th time slot. The probabilities of the possible outcomes of the $n$-th die are given by $\Pr\{X_n^M(i)=m\}=p_n^{(m)},\,\, 1\leq m\leq M$. Therefore, we obtain the optimal mode selection policy given in Theorem \ref{Prot} by substituting the die probabilities in this appendix by the corresponding $p_n^{(m)}$. This completes the proof.

%%%%%%%%%%%%%%%%%%%%%%%%%%%%%%%%%%%%%%%%%%%%%% End One Column Proof %%%%%%%%%%%%%%%%%%%%%%%%%%%%%%%%%%%%%%%%

}{%
  % TwoColumn

%%%%%%%%%%%%%%%%%%%%%%%%%%%%%%%%%%%%%%%%% Begining One Column Proof %%%%%%%%%%%%%%%%%%%%%%%%%%%%%%%%%%%%%%%%

\appendices

\section{Proof of Theorem \ref{Prot}}

Because of space constraints, a detailed proof of Theorem \ref{Prot} is provided in \cite[Appendix A]{ICCArxiv} which is an extended version of the submitted paper.
Herein, we only highlight the
main steps in obtaining the protocol. We first relax the binary condition for $q_k(i)\in\{0,1\}$ to $0\leq q_k(i)\leq 1$. Moreover, since the relaxed
optimization problem is a linear programming problem, an optimal solution is achieved at the boundary of the constraints, i.e., by binary $q_k(i)$. Therefore, the binary relaxation does not affect the maximum achievable sum throughput, please see \cite[Appendix A]{ICCArxiv} for detailed proof. In the following, we investigate the Karush-Kuhn-Tucker (KKT) conditions for the relaxed optimization problem. We note that the relaxed optimization problem is a linear program and thus, the KKT conditions are both necessary and sufficient conditions for optimality.

To simplify the usage of the KKT conditions, we change the maximization of $\bar{R}_{12}+\bar{R}_{21}$ to the
minimization of $-(\bar{R}_{1r}+\bar{R}_{2r})$ since due to constraints $\mathrm{C1}$ and $\mathrm{C2}$ in (\ref{Prob}), $\bar{R}_{12}=\bar{R}_{r2}=\bar{R}_{1r}$ and $\bar{R}_{21}=\bar{R}_{r1}=\bar{R}_{2r}$ must hold. Moreover, without loss of generality, we write all inequalities
in the form $f(x)\leq 0$. The Lagrangian function for the relaxed optimization problem is given by
%%%%%%%%%%%%%%%%%%%%%%%%%%%%%%%%%%%%%%%%%%%%%%%% EQUATION %%%%%%%%%%%%%%%%%%%%%%%%%%%%%%%%%%%%%%%%%
\begin{IEEEeqnarray}{lll}\label{KKTFunction}
   \mathcal{L}  = 
 &-(\bar{R}_{1r}+\bar{R}_{2r}) + \mu_1(\bar{R}_{1r}-\bar{R}_{r2}) + \mu_2(\bar{R}_{2r}-\bar{R}_{r1})  \nonumber \\
   &+ \mathop \sum \limits_{i = 1}^N \lambda \left( i \right)\left( {\mathop \sum \limits_{k = 1}^7 {q_k}\left( i \right) - 1} \right)\nonumber \\
&
+ \mathop \sum \limits_{i = 1}^N \mathop \sum \limits_{k = 1}^7 {\alpha _k}\left( i \right)\left( {{q_k}\left( i \right) - 1} \right)    - \mathop \sum \limits_{i = 1}^N \mathop \sum \limits_{k = 1}^7 {\beta _k}\left( i \right){q_k}\left( i \right)  \IEEEeqnarraynumspace \IEEEyesnumber
\end{IEEEeqnarray}
%%%%%%%%%%%%%%%%%%%%%%%%%%%%%%%%%%%%%%%%%%%%%%%%%%%%%%%%%%%%%%%%%%%%%%%%%%%%%%%%%%%%%%%%%%%%%%%%%%%%%
where $\mu_1,\mu_2$, and $\lambda(i)$ are the Lagrange multipliers corresponding to constraints $\mathrm{C1,C2}$, and $\mathrm{C3}$ in (\ref{Prob}), respectively. Furthermore, $\alpha_k(i)$ and $\beta_k(i)$ are the Lagrange multipliers corresponding to the upper and lower limits of constraint $0\leq q_k(i)\leq 1$, respectively.

In order to determine the optimal selection policy, $q_k^*(i)$, we have to calculate the derivatives of the Lagrangian function in (\ref{KKTFunction}) with respect to $q_k(i)$. This leads to
%%%%%%%%%%%%%%%%%%%%%%%%%%%%%%%%%%%%%%%%%%%%%%%% EQUATION %%%%%%%%%%%%%%%%%%%%%%%%%%%%%%%%%%%%%%%%%
\begin{IEEEeqnarray}{lll}\label{Stationary Mode}
    \frac{\partial\mathcal{L}}{\partial q_1(i)} \Equal - \frac{1}{N}(1\Minus\mu_1)O_1(i)R_{0}\Add \lambda(i)\Add \alpha_1(i)\Minus \beta_1(i) \nonumber  \Equal 0\, \IEEEeqnarraynumspace \IEEEyesnumber \IEEEyessubnumber \\
    \frac{\partial\mathcal{L}}{\partial q_2(i)} \Equal -\frac{1}{N}(1\Minus\mu_2)O_2(i)R_{0}\Add\lambda(i)\Add\alpha_2(i)\Minus\beta_2(i) \nonumber  \Equal 0 \IEEEyessubnumber   \\
    \frac{\partial\mathcal{L}}{\partial q_3(i)} \Equal -\frac{1}{N}(2\Minus \hspace{-0.5mm} \mu_1 \hspace{-0.5mm}\Minus \hspace{-0.5mm}\mu_2)O_3(i)R_{0} \hspace{-0.5mm}\Add\hspace{-0.5mm} \lambda(i) \hspace{-0.5mm}\Add \hspace{-0.5mm} \alpha_3(i)\Minus\beta_3(i) \hspace{-0.5mm}\Equal\hspace{-0.5mm} 0 \,\qquad  \IEEEyessubnumber \\
    \frac{\partial\mathcal{L}}{\partial q_4(i)} \Equal - \frac{1}{N}\mu_2 O_4(i)R_{0}\Add\lambda(i)\Add\alpha_4(i)\Minus \beta_4(i)\Equal 0 \qquad \, \IEEEyessubnumber \\
    \frac{\partial\mathcal{L}}{\partial q_5(i)} \Equal -\frac{1}{N}\mu_1 O_5(i) R_{0}\Add\lambda(i)\Add\alpha_5(i)\Minus\beta_5(i) \Equal 0 \qquad \, \IEEEyessubnumber \\
    \frac{\partial\mathcal{L}}{\partial q_6(i)} \Equal -\frac{1}{N}(\mu_1\Add\mu_2) O_6(i)R_{0}\Add\lambda(i)\nonumber +\alpha_6(i)\Minus\beta_6(i)\Equal 0 \IEEEyessubnumber \\
    \frac{\partial\mathcal{L}}{\partial q_7(i)} \Equal \lambda(i)\nonumber +\alpha_7(i)\Minus\beta_7(i)\Equal 0. \IEEEyessubnumber
\end{IEEEeqnarray}
%%%%%%%%%%%%%%%%%%%%%%%%%%%%%%%%%%%%%%%%%%%%%%%%%%%%%%%%%%%%%%%%%%%%%%%%%%%%%%%%%%%%%%%%%%%%%%%%%%%%%
Without loss of generality, we first obtain the necessary condition for $q_1^*(i)=1$ and then generalize the result 
to $q_k^*(i)=1,\,\,k=2,\dots,7$. If $q_k^*(i)=1$, from constraint $\mathrm{C3}$ in (\ref{Prob}), the other
selection variables are zero, i.e., $q_k^*(i)=0,\,\,k=2,...,7$. Furthermore, from the complementary slackness condition,
we obtain that if an inequality is inactive, the respective Lagrange multiplier must be zero, i.e., $\alpha_k(i)=0,\,\,k = 2,...,7$ and $ \beta_1(i)=0$ have to hold. By substituting these values into (\ref{Stationary Mode}), we obtain
%%%%%%%%%%%%%%%%%%%%%%%%%%%%%%%%%%%%%%%%%%%%%%%% EQUATION %%%%%%%%%%%%%%%%%%%%%%%%%%%%%%%%%%%%%%%%%
\begin{IEEEeqnarray}{lll}\label{MetApp}
    \lambda(i)+\alpha_1(i) = (1-\mu_1)O_1(i)R_{0}  \triangleq \Lambda_1(i) \IEEEyesnumber\IEEEyessubnumber  \\
    \lambda(i)-\beta_2(i) =  (1-\mu_2)O_2(i)R_{0} \triangleq \Lambda_2(i)\IEEEyessubnumber  \\
   \lambda(i)-\beta_3(i) =  (2\Minus \mu_1\Minus \mu_2)O_3(i)R_{0} \triangleq  \Lambda_3(i) \quad\,\,\,\, \IEEEyessubnumber \\
   \lambda(i)-\beta_4(i)  = \mu_2 O_4(i)R_{0} \triangleq \Lambda_4(i) \IEEEyessubnumber\\
    \lambda(i)-\beta_5(i) = \mu_1 O_5(i)R_{0} \triangleq \Lambda_5(i) \IEEEyessubnumber\\
    \lambda(i)-\beta_6(i) = (\mu_1\Add\mu_2) O_6(i)R_{0} \triangleq \Lambda_6(i),\qquad\IEEEyessubnumber \\
     \lambda(i)-\beta_7(i) = 0 \triangleq \Lambda_7(i),\qquad\IEEEyessubnumber
\end{IEEEeqnarray}
%%%%%%%%%%%%%%%%%%%%%%%%%%%%%%%%%%%%%%%%%%%%%%%%%%%%%%%%%%%%%%%%%%%%%%%%%%%%%%%%%%%%%%%%%%%%%%%%%%%%%
where $\Lambda_k(i)$ is referred to as selection metric. By subtracting (\ref{MetApp}b)-(\ref{MetApp}g) from (\ref{MetApp}a), we obtain
%%%%%%%%%%%%%%%%%%%%%%%%%%%%%%%%%%%%%%%%%%%%%%%% EQUATION %%%%%%%%%%%%%%%%%%%%%%%%%%%%%%%%%%%%%%%%%
\begin{IEEEeqnarray}{rCl}\label{eq_2_1}
    \Lambda_1(i) - \Lambda_k(i) = \alpha_1(i)+\beta_k(i), \quad k=2,\dots ,7. \IEEEyesnumber
\end{IEEEeqnarray}
%%%%%%%%%%%%%%%%%%%%%%%%%%%%%%%%%%%%%%%%%%%%%%%%%%%%%%%%%%%%%%%%%%%%%%%%%%%%%%%%%%%%%%%%%%%%%%%%%%%%%
Moreover, the dual feasibility conditions  for Lagrange multipliers imply $\alpha_k(i),\beta_k(i)\geq 0$. By inserting $\alpha_k(i),\beta_k(i)\geq 0$ in (\ref{eq_2_1}), we obtain a necessary condition for $q_1^*(i)=1$ as
%%%%%%%%%%%%%%%%%%%%%%%%%%%%%%%%%%%%%%%%%%%%%%%% EQUATION %%%%%%%%%%%%%%%%%%%%%%%%%%%%%%%%%%%%%%%%%
\begin{IEEEeqnarray}{ccc}
    \Lambda_1(i) \geq \max \left \{ \Lambda_2(i), \Lambda_3(i), \Lambda_4(i), \Lambda_5(i), \Lambda_6(i),\Lambda_7(i) \right \}. \quad\IEEEyesnumber
\end{IEEEeqnarray}
%%%%%%%%%%%%%%%%%%%%%%%%%%%%%%%%%%%%%%%%%%%%%%%%%%%%%%%%%%%%%%%%%%%%%%%%%%%%%%%%%%%%%%%%%%%%%%%%%%%%%
Repeating the same procedure for $q_k^*(i)=1,\,\,k=2,\dots,7$, we obtain a necessary condition for selecting transmission mode $\mathcal{M}_{k^*}$ in the $i$-th time slot as 
%%%%%%%%%%%%%%%%%%%%%%%%%%%%%%%%%%%%%%%%%%%%%%%% EQUATION %%%%%%%%%%%%%%%%%%%%%%%%%%%%%%%%%%%%%%%%%
\begin{IEEEeqnarray}{ccc}
    \Lambda_{k^*}(i) \geq {\underset{k\neq k^*}{\max}}\{\Lambda_{k}(i)\}. \quad\IEEEyesnumber
\end{IEEEeqnarray}
%%%%%%%%%%%%%%%%%%%%%%%%%%%%%%%%%%%%%%%%%%%%%%%%%%%%%%%%%%%%%%%%%%%%%%%%%%%%%%%%%%%%%%%%%%%%%%%%%%%%%
Hence, we can conclude that the transmission modes with the largest selection metrics are the candidates for optimal mode selection. 

Now, in order to obtain the optimal transmission mode, we need to answer the following questions: 1) what are the optimal values of selection weights $\mu_1$ and $\mu_2$ in order to calculate the selection metrics, and 2) given $\mu_1$ and $\mu_2$, how should we  select the optimal mode from the modes with identical selection metric values. In \cite[Appendix A]{ICCArxiv}, we proved that  $0\leq \mu_1+\mu_2\leq 1$, $1-2\mu_1\leq \mu_2 \leq 1$, and $1-2\mu_2 \leq \mu_1 \leq 1$ have to hold to fulfill constraints $\mathrm{C1}$ and $\mathrm{C2}$ in (\ref{Prob}). Moreover, the exact values of $\mu_1$ and $\mu_2$ lead to the statistical regions that are given in Table II. We also show in \cite[Appendix A]{ICCArxiv} that for the modes with identical selection metric values, different selection policies that satisfy constraints $\mathrm{C1}$ and $\mathrm{C2}$ in (\ref{Prob}) may lead to the same sum throughput. Therefore, for simplicity, we adopt a probabilistic approach via rolling dice where the die probabilities are chosen to satisfy $\mathrm{C1}$ and $\mathrm{C2}$ in (\ref{Prob}) and are given in Table II. This competes the proof.

%%%%%%%%%%%%%%%%%%%%%%%%%%%%%%%%%%%%%%%%%%%%%% End One Column Proof %%%%%%%%%%%%%%%%%%%%%%%%%%%%%%%%%%%%%%%%

}

%%%%%%%%%%%%%%%%%%%%%%%%%%%%%%%%%%%% Bibliography %%%%%%%%%%%%%%%%%%%%%%%%%%%%%%%%%%%%%%%%%%%%%%%%%%%%%
\bibliographystyle{IEEEtran}
\bibliography{Ref_31_01_2014}
%%%%%%%%%%%%%%%%%%%%%%%%%%%%%%%%%%%%%%%%%%%%%%%%%%%%%%%%%%%%%%%%%%%%%%%%%%%%%%%%%%%%%%%%%%%%%%%%%%%%%

%\end{doublespace}
\end{document}